\documentclass[11pt]{article}
\usepackage[english]{babel}
\addto{\captionsenglish}{}
\usepackage{amsmath,amsthm}
\usepackage{amsfonts}
\usepackage[dvips]{graphics}
\usepackage{graphicx,latexsym,amssymb,amsmath,cancel,color,enumerate,multirow,lscape,bm,rotating,multirow}

\usepackage{booktabs}
\usepackage[svgnames]{xcolor}
\usepackage[pass]{geometry}
\usepackage{pdflscape}
\usepackage{caption}
\usepackage{subcaption}
\usepackage{soul}
\usepackage{bbm}
\usepackage[all]{xy}
\usepackage[round]{natbib}
\usepackage{setspace}
\usepackage[section]{placeins}	
\usepackage{float}	
\usepackage[T1]{fontenc}
\usepackage{lmodern}
\usepackage[nottoc,notlot,notlof]{tocbibind}
\usepackage{hyperref}
\usepackage{esint}
\usepackage{leftidx}
\usepackage{lineno}

\usepackage{epsf}
\usepackage{makeidx}
\usepackage{multicol}
\usepackage{color, colortbl}

\definecolor{beaublue}{rgb}{0.74, 0.83, 0.9}
\definecolor{aliceblue}{rgb}{0.94, 0.97, 1.0}
\definecolor{interpolateblue}{rgb}{0.84, 0.9, 0.95}

\newtheorem{prop}{Proposition}

\theoremstyle{remark}
\newtheorem{remark}{Remark}

\newcommand{\argmin}{\mathop{\mathrm{argmin}}}

\usepackage{graphicx,scalerel,mathtools}

\newcommand{\ind}{{\rm I\hspace{-2.3mm} 1}}

\begin{document}
\title{On inference in high-dimensional regression}
\author{H.~S.~Battey and N.~Reid
\footnote{\,Heather Battey ({\tt h.battey@imperial.ac.uk}). Department of Mathematics, Imperial College London, 180 Queen's Gate, London SW7 2AZ, UK. \newline
{\color{white}sp}		 $^*$\,Nancy Reid ({\tt  nancym.reid@utoronto.ca}). Department of Statistical Sciences, University of Toronto, 700 University Ave, 9th Floor, Toronto, Ontario M5G 1Z5, Canada.}
}
\date{\today}
\maketitle

\begin{abstract}
This paper develops an approach to inference in a linear regression model when the number of potential explanatory variables is larger than the sample size. The approach treats each regression coefficient in turn as the interest parameter, the remaining coefficients being nuisance parameters, and seeks an optimal interest-respecting transformation, inducing sparsity on the relevant blocks of the notional Fisher information matrix. The induced sparsity is exploited through a marginal least squares analysis for each variable, as in a factorial experiment, thereby avoiding penalization. One parameterization of the problem is found to be particularly convenient, both computationally and mathematically. In particular, it permits an analytic solution to the optimal transformation problem, facilitating theoretical analysis and comparison to other work.  In contrast to regularized regression such as the lasso and its extensions, neither adjustment for selection nor rescaling of the explanatory variables is needed, ensuring the physical interpretation of regression coefficients is retained. Recommended usage is within a broader set of inferential statements, so as to reflect uncertainty over the model as well as over the parameters. The considerations involved in extending the work to other regression models are briefly discussed.
\end{abstract}

\bigskip{\noindent{\bf Some key words:} conditionality; confidence sets of models; factorial contrasts; fixed design; inducement of sparsity; nuisance parameters; parameter orthogonalization.}

\maketitle

\section{Introduction}\label{secIntro} 

We consider inference for the coefficient parameters of a high-dimensional linear regression model in which the number of potential explanatory variables $p$ is larger than the sample size $n$. In this context, the debiased lasso \citep{ZZ2014, vdG2014} and decorrelated score \citep{NingLiu2017} require two sparsity assumptions: a natural one on the parameter vector, and a less natural one on the inverse Fisher information matrix. The key contribution of the present paper is an approach that induces sparsity on relevant blocks of the notional Fisher information matrix, removing the second assumption without affecting interpretation of the parameter of interest. The induced sparsity is exploited through a marginal least squares analysis for each variable, as in a factorial experiment. Since regularized regression is not used, there is no need to standardize the columns of the covariate matrix to have unit length. Thus, the physical interpretation of coefficients is retained.

Primary inspiration for the proposal came from the analysis of matched comparison problems, in which it is sometimes feasible to eliminate a large number of nuisance parameters by simple operations on the sample space. There are also connections to \cite{CR1987} as discussed in \S \ref{secCS}. Section \ref{secLiterature} provides a more detailed comparison to the debiased lasso and related proposals.

The remainder of the present section focusses on the broader inferential strategy. Usual usage of the debiased lasso and its extensions, as described for example in \cite{BuhlmannAR}, involves hypothesis tests for all the parameters followed by a correction for multiple testing. The conclusion is typically that the null hypothesis of no effect is rejected for a very small number of variables; the joint explanatory power of sets of variables tends not to be assessed. The present paper instead proposes confidence intervals for the regression coefficient parameters as an adjunct to confidence sets of models \citep{CB2017,BC2018}, as detailed in \S \ref{secConfModels}. The latter papers emphasized the construction of statistically indistinguishable sets of variables based on their fit to the data, but did not consider inferential statements about the individual parameters or their associated compatibility with the models in the confidence set. This strategy is illustrated on a set of gene-expression data from \cite{BuhlmannAR}.

 Several quite different approaches to inference in high-dimensional regression have been proposed. In  \cite{Lockhart2014}, \cite{Lee2016}, and \cite{TT2018} inferential statements are corrected for variable selection by conditioning on the region of the sample space that led to the selection being made. Sample-splitting is a simple alternative that avoids detailed characterization of the selection event. The efficiency loss of sample-splitting was calculated in a simple example by \cite{Cox1975} and several more elaborate splitting strategies have been proposed in \cite{RasinesYoung2021} and \citet{Ramdas2021}. For a different perspective see \cite{Witten2021}. 
The approach advocated by \cite{Berk2013}, and further discussed in \cite{Leeb2015}, aims instead  for inferential statements that are universally valid under all possible model selection procedures; this leads to highly conservative confidence regions.

In contrast with much of the earlier literature on this topic, the analysis of \S \ref{secTheory} is conditional on $X$, as would be dictated by ancillarity  when $p<n$. In addition to  ancillarity, there are substantive reasons for preferring a conditional analysis. In observational studies $X$, although random, is frequently realized in advance of the response variable $Y$, in which case it may be possible to interpret the effect of $X$ on $Y$ as causal. For example $X$ may be information on the genome or epi-genome, and $Y$ a measure of some condition expected to be caused by changes in $X$. However, even in cases where the values of $(Y_i, X_i)_{i=1}^n$ are realized contemporaneously, it is appropriate to condition on $X$ for studying its association  with the outcome variable $Y$. Otherwise it would be possible that an inference justified on the basis of its unconditional properties performed poorly for any given sample of $X$.

\section{Approximate orthogonalization}\label{secCS}

Let $x_i$ be a vector of measurements on $p$ covariates for individual $i$. Outcomes are assumed to be generated from a linear regression model 
\[
Y_{i}=x_{i}^T\beta + \varepsilon_i, \quad i=1,\ldots n,
\]
where $(\varepsilon_i)_{i=1}^n$ are independent with mean  zero  and variance $\tau>0$, and $\beta = (\beta_1, \ldots, \beta_p)^{T}$ is of dimension $p\gg n$, but satisfying the sparsity condition $\|\beta\|_0=\sum_{u=1}^{p}\ind\{\beta_u \neq 0\}=s \ll p$. The set of signal indices is $\mathcal{S}=\{v:\beta_v \neq 0\}$. The arbitrary designated interest parameter is $\beta_v$, so that the vector $\beta_{-v}$ of nuisance parameters has elements $\{\beta_1,\ldots,\beta_p\}\backslash \beta_v$. Let $Y=(Y_1,\ldots, Y_n)^{T}$ and let $X$ be the matrix whose $i$th row is $x_i^T$. The column of $X$ corresponding to the interest parameter $\beta_v$ is written $x_{v}=(x_{1,v},\ldots,x_{n,v})^{T}$, where indices $u,v,w,\ldots$ from the end of the alphabet distinguish arbitrary columns of $X$ from arbitrary transposed rows. The remaining columns are those of the $n\times (p-1)$ dimensional matrix, $X_{-v}$, whose $i$th row is $x^T_{i,-v}$. On writing $\varepsilon = (\varepsilon_1,\ldots,\varepsilon_n)^T$, the linear regression model is
\begin{equation}\label{eqLinearReg}
Y= X\beta +\varepsilon = x_{v}\beta_v +X_{-v}\beta_{-v} + \varepsilon.
\end{equation}

If the columns in $X_{-v}$ are orthogonal to $x_v$ then a simple regression of $Y$ on $x_v$ estimates $\beta_v$ without bias, this being the motivation for factorial experiments and more elaborate experimental designs. In fact only the columns corresponding to signal variables need be orthogonal to $x_v$, but it is not known which columns these may be.

Since the parameter $\beta$ in \eqref{eqLinearReg} is unchanged by premultiplication of both sides by an $n\times n$ matrix, for each interest parameter $\beta_v$ a matrix $A^v$ is sought such that the columns $\widetilde{x}_{w}^v$ of the transformed covariate matrix $\widetilde{X}^v=A^v X$ are as orthogonal as possible to $\widetilde{x}_v^{v}$. The parameter of interest $\beta_v$ is then estimated by simple linear regression of $\widetilde{Y}^v = A^v Y$ on $\widetilde {x}_v^v$. The associated least squares estimator is
\begin{equation}\label{eqTildePsi}
\widetilde{\beta}_v=(\widetilde{x}_{v}^{v T}\widetilde{x}_{v}^{v})^{-1}\widetilde{x}_{v}^{v T}\widetilde{Y}^{v}= (x_{v}^{T}A^{v T}A^v x_{v})^{-1}x_{v}^{T}A^{v T} A^{v} Y.
\end{equation}
Under model (\ref{eqLinearReg})
\begin{eqnarray}\label{eqBiasVar}
\mathbb{E}(\widetilde{\beta}_v) & = & \beta_v + \textstyle{\sum_{w\in \mathcal{S}}}  \vartheta_{w} \, \beta_w =: \beta_v + b_v, \\
\nonumber \text{var}(\widetilde{\beta}_v) & = &  (\widetilde{x}_{v}^{v T}\widetilde{x}_{v}^v)^{-1}\widetilde{x}_{v}^{v T} A^{v}\mathbb{E}(\varepsilon \varepsilon^{T}) A^{v T} 	\widetilde{x}_{v}^{v} (\widetilde{x}_{v}^{v T}\widetilde{x}_{v}^v)^{-1} =: \tau \sigma_{vv},
\end{eqnarray}
where
\begin{eqnarray}\label{eqBiasingTerms}
\vartheta_{w}=\vartheta_{w}(A^v) &=& (\widetilde{x}_{v}^{v T}\widetilde{x}_{v}^v)^{-1}\widetilde{x}_{v}^{v T} \widetilde{x}_{w}^{v}, \\ 
\nonumber \sigma_{vv} = \sigma_{vv}(A^v) &=& (\widetilde{x}_{v}^{v T}\widetilde{x}_{v}^v)^{-2}\widetilde{x}_{v}^{v T} A^{v}A^{v T}\widetilde{x}_{v}^{v},
\end{eqnarray}
and by the Cauchy-Schwarz inequality, $b_v^2\leq \|\beta_{-v}\|_2^2 \sum_{w\in \mathcal{S}} \vartheta^2_{w}$. Since $\|\beta_{-v}\|_{2}$ and $\mathcal{S}$ are unknown, this suggests choosing $A^v$ to minimize $\sigma_{vv}(A^v)+\sum_{w\neq v}\vartheta_{w}^2(A^v)$, an upper bound on the observable components of the mean squared error. The relevant objects in terms of $q_v=A^{v T} A^v x_{v}$ are
\begin{eqnarray}\label{eqqPsi}
\widetilde{\beta}_v & = & (q_{v}^{T}x_{v})^{-1}q_{v}^{T} Y, \\
\nonumber \sigma_{vv}  &=& (q_{v}^{T}x_{v})^{-2} (q_{v}^{T}q_{v}), \\
\nonumber \vartheta_{w}^2  &=& (q_{v}^{T}x_{v})^{-2} (q_{v}^{T}x_{w})^2.
\end{eqnarray}
On noting that
\[
(q_{v}^{T}x_{w})^2=q_{v}^{T}x_{w}x_{w}^T q_{v},
\] 
the sum of the variance $\tau \sigma_{vv}(q_{v})$ and squared potential biasing terms $\textstyle{\sum_{w\neq v}} \vartheta_{w}^2(q_{v})$ is minimized when $q_v$ solves the simple unconstrained optimization problem 
\[
\argmin_{q\in \mathbb{R}^n} \; (q^{T}x_{v})^{-2}q^{T}(I_n+\textstyle{\sum_{w\neq v}} x_{w}x_{w}^{T})q.
\]
This estimator, which weights squared bias and variance equally, is a natural choice. However, a generalization is helpful for gaining geometric insight into the problem and for comparison with other work. Thus define $q_v$ by
\begin{equation}\label{eqOptimalq}
q_{v} \in \mathcal{Q}_{v}^{(\delta)} = \argmin_{q\in \mathbb{R}^n} \; (q^{T}x_{v})^{-2}q^{T}(\delta I_n+X_{-v}X_{-v}^{T})q,
\end{equation}
for some fixed $\delta>0$ to be chosen. In \S \ref{secTheory}, such solutions will be shown to exist in closed form under a condition on $X$ defined below at Proposition \ref{propq}. If $q_v\in \mathcal{Q}_v^{(\delta)}$, then any non-zero scalar multiple of $q_v$ is also in $\mathcal{Q}_{v}^{(\delta)}$, as can be seen from the form of the objective function. Hereafter, $q_v$ refers to an arbitrary solution to the minimization problem of equation \eqref{eqOptimalq} unless otherwise specified.

The connection to parameter orthogonalization \citep{CR1987} is that, for Gaussian linear regression, both approaches seek to induce sparsity on the relevant row and column of the Fisher information matrix. With $\beta_v$ identified as the interest parameter, a so-called interest-respecting orthogonal parameterization of model (\ref{eqLinearReg}) is $(\beta_v,\eta)$, where
\begin{equation}\label{eqEta}
\eta := X_{-v}^Tx_{v}\beta_v + X_{-v}^TX_{-v}\beta_{-v},
\end{equation}
or alternatively, if $(X_{-v}^T X_{-v})^{-1}$ exists, $(\beta_v,\phi)$, where
\begin{equation}\label{eqPhi}
\phi := (X_{-v}^T X_{-v})^{-1}\eta =\beta_{-v} +  (X_{-v}^T X_{-v})^{-1}X^{T}_{-v} x_v \beta_v,
\end{equation}
as orthogonal parameters are only defined up to linear transformation.  Expressing (\ref{eqLinearReg}) in either of parameterizations \eqref{eqEta} or \eqref{eqPhi} requires $p<n$.  This suggests an alternative  to \eqref{eqPhi} in which $(X_{-v}^T X_{-v})^{-1}$  is replaced by a regularized inverse $(\delta I_{p} + X_{-v}^T X_{-v})^{-1}$, an approach discussed briefly in \S \ref{secDiscuss}.

Section \ref{secTheory} derives some theoretical aspects of the above approximate orthogonalization procedure, particularly the interpretation of $q_v$, which clarifies the relationship between the present proposal and that of \cite{ZZ2014}. Proofs of the main results are given in Appendix \ref{secProofs}.

\begin{remark}
For any two parameters $\beta_u$ and $\beta_v$, the covariance between $\widetilde{\beta}_u$ and $\widetilde{\beta}_v$ constructed according to equations \eqref{eqTildePsi} and \eqref{eqOptimalq}, is
\begin{equation}\label{eqCovariance}
\text{cov}(\widetilde{\beta}_u,\widetilde{\beta}_v) = \tau (q_{u}^{T}x_{u})^{-1} q_{u}^{T}q_{v} (q_{v}^T x_{v})^{-1} =:\tau \sigma_{uv},
\end{equation}
which is known up to $\tau$. 
\end{remark}

\begin{remark} 
	There is a simple transformation of $X$, $\mathring{X}^v$, say, whose columns are exactly orthogonal to a given one, $\mathring{x}^v_v$ say. This is $\mathring{X}^v = \mathring{P}^{v}X + C^v$ where $\mathring{P}^{v}=I_{n}-(x_v^T x_v)^{-1}x_{v}x_v^T$ projects the columns of $X$ into the subspace of $\mathbb{R}^n$ orthogonal to $x_v$, and $C^v$ is a matrix of zeros apart from the column $c^v_v=\mathring{x}^v_v=x_v$, which replaces the single column of zeros produced by $\mathring{P}^v X$. Since this transformation to $\mathring{X}^v$ is not of the form $A^v X$, it is not interest-respecting. An interest-respecting version could be obtained by projection of $\mathring{X}^v$, in a suitable matrix norm, into the space of matrices of the form $AX$, where $A\in \mathbb{R}^{n\times n}$. This is however much less convenient than \eqref{eqOptimalq}, both computationally and theoretically.
	\end{remark}

A nominal $\alpha$-level confidence interval for $\beta_v$ is
\begin{equation}\label{eqCSPsi}
\widetilde{\mathcal{C}}_{v}(\alpha) = [\,\widetilde{\beta}_v-z_{1-\alpha/2}(\tau \sigma_{vv})^{1/2}, \widetilde{\beta}_v+z_{1-\alpha/2}(\tau \sigma_{vv})^{1/2}\,],
\end{equation}
where $\Phi(z_{1-\alpha}) = \alpha$. As shown in \S \ref{secTheory}, the confidence interval $\eqref{eqCSPsi}$ is valid as $n\rightarrow \infty$, to the extent that $q_v$ is successful in orthogonalizing $\widetilde{x}^v_{v}$ to $\widetilde{x}^{v}_{w}$ for all $w \in \mathcal{S}$. The qualification here is because the objective function in \eqref{eqOptimalq} does not ensure $\widetilde{x}^v_{v}$ is exactly orthogonal to $\widetilde{x}^{v}_{w}$ for all $w\in \mathcal{S}$, only that the cumulative non-orthogonality to $\widetilde{x}^{v}_{w}$ over $w\neq v$ is small. For some $v$ this non-orthogonality will concentrate relatively more on the unknown signal indices, resulting in larger bias for those parameters. However, as shown in Proposition \ref{propBiasb}, this bias decays rapidly with the sample size.

One could in principle attempt to identify positions where the signal variables are most likely to be and adjust the objective function \eqref{eqOptimalq} accordingly, a suggestion to which we return in \S \ref{secDiscuss}. A variant of this idea has been proposed recently by \cite{Ziwei2021}. Since their procedure is based on the debiased lasso \citep{ZZ2014, vdG2014} the resulting inference is not scale-invariant, and the theoretical guarantees entail a similar condition of sparsity of the inverse Fisher information matrix, which in their random design setting corresponds to sparsity of the inverse covariance matrix of the covariates.

\section{Inferential properties}\label{secTheory}

In setting up the objective function
\begin{equation}\label{objFunction}
m(q) = (q^{T}x_{v})^{-2}q^{T}(\delta I_n+X_{-v}X_{-v}^T)q
\end{equation}
in equation \eqref{eqOptimalq}, $\delta> 0$ was introduced to enable comparison to other work. However, the procedure is remarkably robust to this choice, a phenomenon discussed in further detail following Proposition \ref{propBiasb}. 

Let $(\delta I_{n} + X_{-v} X_{-v}^T)=M_{\delta}$, and define%
\begin{equation}\label{eqLdelta}
L_{\delta} 
= M_{\delta} - \{x_v^T M_{\delta}^{-1}x_v\}^{-1}x_v x_v^T.
\end{equation}
Proposition \ref{propq} specifies the set of solutions $\mathcal{Q}_v^{(\delta)}$ from \eqref{eqOptimalq} in closed form.

\begin{prop}\label{propq}
Suppose $\delta>0$  is such that the eigenvalues of $L_{\delta}$ are non-negative. Then any $q_v$ of the form
\begin{equation}\label{eqExpressionq}
	q_v = a (\delta I_n + X_{-v}X_{-v}^T)^{-1}x_v
\end{equation}
is a minimizer of \eqref{objFunction}, where $a$ is any non-zero real number. If the eigenvalue condition is violated, then such a $q_v$ is a saddlepoint of \eqref{objFunction}.
	\end{prop}

Modulo multiplication by a positive scalar, $L_{\delta}$ is the matrix field $\nabla \nabla^T m$ 
evaluated at any $q_v$ of the form \eqref{eqExpressionq}. Let $z =\{x_{v}^T M_\delta^{-1} x_v\}^{-1/2}x_v$ so that $L_{\delta}=M_{\delta}-z z^T$. Then
\[
\text{det}(L_{\delta})=(1-z^{T}M_\delta^{-1}z)\text{det}(M_\delta) =0,
\]
where the first equality is the so-called matrix determinant lemma for determinants of rank one perturbations and the last equality follows as $z^{T}M_\delta^{-1} z=1$.
Thus, the matrix $L_{\delta}$ has at least one zero eigenvalue, showing that, if the condition of Proposition \ref{propq} is satisfied, $m$ is locally weakly convex at any $q_v$ of the form \eqref{eqExpressionq} but not strictly convex \citep[e.g.][]{BoydVand}. Let
\begin{equation}\label{eqDefP}
P^{v}(a,\delta) = a (\delta I_n + X_{-v}X_{-v}^T)^{-1}.
\end{equation}
so that $q_v = P^{v}(a,\delta)x_v$. Although any choice of $a$ yields a minimizer of \eqref{objFunction}, the choice $a=\delta$ improves interpretability of \eqref{eqDefP} in view of equation \eqref{eqGeneralP} of Proposition \ref{propbias}. It is also of some theoretical value to consider $a=\delta\rightarrow 0^{+}$.

\begin{prop}\label{propbias}
Set $a=\delta$ and define $X_{-v}^{+}$ to be the Moore-Penrose pseudo-inverse of $X_{-v}$ given in \eqref{eqMP} of Appendix \ref{secProofBias}. Then
\begin{equation}\label{eqGeneralP}
P^v(\delta, \delta)=I_{n} - X_{-v}(\delta I_{p-1}+X_{-v}^{T}X_{-v})^{-1}X_{-v}^{T}
\end{equation}
and
\begin{equation}\label{eqLimitingP}
P^v := \lim_{\delta \rightarrow 0^{+}}P^{v}(\delta,\delta) = I_n -X_{-v}X_{-v}^{+}
\end{equation}
is an orthogonal projection operator that maps any vector in $\mathbb{R}^n$ into the left null space of $X_{-v}$ or, equivalently, the kernel of $X_{-v}^T$, $\emph{\text{ker}}(X_{-v}^T)=\{u\in\mathbb{R}^n: X_{-v}^T u =0\}\subset \mathbb{R}^n$.
\end{prop}

This limiting case only serves to supply insight and is of low practical relevance when $p>n$. When $p<n$, $X_{-v}^{+}=(X_{-v}^T X_{-v})^{-1}X_{-v}^T$ so that $P^v=I_{n} - X_{-v}(X_{-v}^T X_{-v})^{-1}X_{-v}^T$ and \eqref{eqGeneralP} can be viewed as a ridge-regularized version of this projection \citep{HK1970}. See \S \ref{secLiterature} for further discussion.

From \eqref{eqqPsi}, \eqref{eqOptimalq}, and Proposition \ref{propAsympS} below, $\tau^{-1/2}S_n (\beta_v)$ is an asymptotically pivotal quantity, where
 \begin{equation}\label{eqSn}
S_n (\beta_v) :=	\frac{(q_{v}^{T}x_v)(\widetilde{\beta}_v - \beta_v - b_v)}{( q_v^{T} q_{v})^{1/2}}.
\end{equation}
However the confidence intervals in \eqref{eqCSPsi} ignore the bias term $ (q_v^Tx_v) b_v/(\tau q_v^T q_v )^{1/2}$. 
While, for sufficiently small $\delta$, $q_v$ effectively minimizes the observable part of an upper bound on $b_v^2$, it is inevitable that $|b_v|$ will be larger for some components $\beta_v$ than for others, due to lack of knowledge of $\mathcal{S}$. Proposition \ref{propBiasb} establishes the behaviour of the bias $b_v$ in \eqref{eqBiasVar} and the quantity $ (q_v^Tx_v) b_v/(\tau q_v^T q_v )^{1/2}$ as a function of the sample size. The simulations in \S \ref{secFactorial} are consistent with these conclusions.

By Proposition \ref{propbias}, $P^v(\delta,\delta)z$ is a regularized projection of $z\in\mathbb{R}^n$ into the kernel of $X_{-v}^T$, so by construction the elements of $X_{-v}^T P^v(\delta,\delta)z$ are small and bounded away from zero provided that $\delta$ is. To avoid explicitly quantifying the dependence on $n$, $p$ and $\delta$, write $X_{-v}^T P^v(\delta,\delta)x_v \asymp h(n,p,\delta)$.

\begin{prop}\label{propBiasb} Let $\delta>0$ be bounded away from zero. Provided that the non-zero elements of $\beta_{-v}$ are bounded, the bias terms in \eqref{eqSn} satisfy 
	\[
	b_v = O(s/n), \quad\ \text{and} \quad \frac{(q_v^T x_v) b_v}{(\tau q_v^T q_v)^{1/2}} = O \biggl(\frac{s}{n^{1/2}}\cdot \frac{h(n,p,\delta)^{1/2}}{\delta^{1/2}} \biggr).
	\]
\end{prop}

Proposition \ref{propBiasb} suggests that one sensible choice of $\delta$ would solve the fixed point equation $\delta\asymp h(n,p,\delta)$, so that 
\[
\frac{(q_v^T x_v) b_v}{(\tau q_v^T q_v)^{1/2}} = O\biggl(\frac{s}{n^{1/2}}\biggr).
\]
We have found empirically that the stability of the conclusions to the choice of $\delta$ is remarkably high. Such stability can always be checked using the data at hand.

In  applying Proposition \ref{propBiasb}, confidence intervals are based on the  asymptotically pivotal quantity $V_n^{-1/2}U_n(\beta_v)$, where $V_n$ is an estimator of $\tau$ discussed below and 
\begin{equation}\label{eqUn}
U_{n}(\beta_v):=	\frac{(q_{v}^{T}x_v)(\widetilde{\beta}_v - \beta_v)}{ (q_v^{T} q_{v})^{1/2}}. 
\end{equation}
The validity of this is established   via Proposition \ref{propBiasb}, combined with the limiting distribution of $S_n(\beta_v)$ derived below. We also see that, for fixed sample size, the bias plays a relatively larger role for components whose associated variances are smaller. This is visually apparent from Figure \ref{figSimulations} of \S \ref{secFactorial}.

\begin{prop}\label{propAsympS}
	Suppose that $(\varepsilon_i)_{i=1}^n$ are  independent mean zero random variables of variance $\tau$ and bounded moments of order $2+\gamma$ for some $\gamma>0$. Then
	\[
	\sup_{z\in \mathbb{R}}|\text{\emph{pr}}\{\tau^{-1/2}S_n(\beta_v)\leq z\} -\Phi(z)| \rightarrow 0, \text{ as } n\rightarrow \infty,
	\]
	where $\Phi$ is the standard normal cumulative distribution function. If the above condition holds with $\gamma= 1$,
	\begin{equation}\label{eqBerryEsseen}
	\sup_{z\in \mathbb{R}}|\text{\emph{pr}}\{\tau^{-1/2}S_n(\beta_v)\leq z\} -\Phi(z)| \leq C e_n,
	\end{equation}
	where $C\in(0.4097, 0.5606]$,  $(q_{j,v})_{j=1}^n$ are the elements of  $q_v$, and 
	\[
	e_n = \frac{\sum_{j=1}^{n}|q_{j,v}|^3 \mathbb{E}|\varepsilon_j|^3}{(n\tau)^{3/2} (q_v^T q_v)^{3/4}}.
	\]
\end{prop}

With any consistent estimator $V_n$ of the error variance, the limiting normal approximation to the Studentized statistic $V_{n}^{-1/2}S_n(\beta_v)$ remains valid by Slutsky's theorem. {Proposition \ref{propStudentized0} provides more precise quantification.

\begin{prop}\label{propStudentized0}
Suppose that the conditions of Proposition \ref{propAsympS} hold with $\gamma= 1$ and let $V_n$ be a consistent estimator of $\tau$. On writing
\begin{equation}\label{eqVarianceConv}
\text{\emph{pr}}(|V_{n}^{-1/2}-\tau^{-1/2}|>t_n) \asymp r_n(t_n),
\end{equation}
where $t_n$ and $r_n(t_n)$ are both decreasing in $n$, the normal approximation to the distribution of $V_n^{-1/2} S_{n}(\beta_v)$ has uniform error of order
	\begin{equation}\label{eqBerryEsseenStudent}
	\sup_{z\in \mathbb{R}}|\text{\emph{pr}}\{V_n^{-1/2}S_n(\beta_v)\leq z\} -\Phi(z)| \leq c\max\{e_n,1-\Phi(g_n),t_n g_n,r_n(t_n)\},
	\end{equation}
	where $c$ is a constant that we do not quantify, $g_n$ is any nondecreasing sequence such that $t_n g_n = o(1)$, and $e_n$ is defined in Proposition \ref{propAsympS}.
	\end{prop}

While it is unclear which choice of $g_n$ delivers the sharpest bound in \eqref{eqBerryEsseenStudent}, a reasonably tight bound is obtained by choosing $g_n$ such that $1-\Phi(g_n)$ is roughly the same order as $e_n$. By Mill's ratio
\[
\frac{1-\Phi(x)}{\phi(x)} \in \Bigl(\frac{1}{x+1/x}, \frac{1}{x}\Bigr), \quad x>0,
\]
where $\phi$ is the standard normal density function. Thus, using the approximation $e_n=g_{n}^{-1}\phi(g_n)$ and solving approximately for $g_n$ using Lagrange's inversion formula \citep[e.g.][p.25--29]{deBruijn}, we obtain
\[
g_n=\Bigl([\log(e_n^{-2})-\log\{\log(e_n^{-2})\}] + O[\log\{\log(e_n^{-2})\}/\log(e_n^{-2})]\Bigr)^{1/2},
\]
which is a very slowly growing function of $n$. Thus, asymptotically, the confidence interval based on the infeasible statistic $V_n^{-1/2}S_{n}(\beta_v)$ has nominal coverage probability $1-\alpha$, and for any fixed $n$ the error in this approximation is roughly of order $\max\{e_n,t_n g_n, r_n(t_n)\}$.}
 
\section{The debiased lasso}\label{secLiterature}

Our proposal is related to inferential methods based on inverting the Karush-Kuhn-Tucker conditions associated with a lasso solution. In one respect, the most general procedure is that of \cite{vdG2014}, which covers penalized likelihood inference. Their debiased lasso estimator, $\widehat{\beta}^\text{d}$ say, reduces essentially to the proposal of \cite{ZZ2014} in the case of the linear model. \cite{ZZ2014} modified the score equation for the least squares estimator  to  $z_v^T(Y-x_v \beta_v)=0$, with $z_v$ to be chosen. By comparison the estimating equation that defines $\widetilde{\beta}_v$ is
\begin{equation}
\widetilde{x}_{v}^{v T}(\widetilde{Y}^v - \widetilde{x}_{v}^{v}\beta_v) = q_v^T(Y - x_{v} \beta_v) = 0, \quad\text{with } q_v = A^{v T}A^v x_v.
\end{equation}
\cite{ZZ2014} suggested a bias correction based an a pilot estimator $\widehat{\beta}_{w}^{\text{(init)}}$ of each nuisance parameter $\beta_{w}$:
\[
\frac{z_v^T Y}{z_v^T x_v} - \sum_{w\neq v}  \frac{ \widehat{\beta}_w^{\text{(init)}} z_v^T x_w}{z_v^T x_v} =  \beta_v + 	\frac{z_v^T \varepsilon}{z_v^T x_v}.
\]
Thus the bias is controlled by
\begin{equation}\label{eqBiasZZ}
\biggl( \frac{\max_{w\neq v}|z_v^{T}x_w|}{|z_v^T x_v|} \biggr) \|\widehat{\beta}_{-v}^{\text{(init)}}-\beta_{-v}\|_{1}
\end{equation}
and \cite{ZZ2014} recommended that $\widehat{\beta}_{-v}^{\text{(init)}}$ be taken as the lasso estimator, so that $\|\widehat{\beta}_{-v}^{\text{(init)}}-\beta_{-v}\|_{1}$ is small with high probability under regularity conditions. They also recommended that $z_{v}$ be chosen using the lasso, motivated by ordinary least squares when $p<n$, where the score equation $z_v^T(Y-x_v \beta_v)=0$ holds for the least squares estimator with
\[
z_v=x_{v}^{\perp}:=(I_n-X_{-v}(X_{-v}^{T}X_{-v})^{-1}X_{-v})x_v,
\] 
i.e., the projection of $x_v$ on the orthogonal complement of the space spanned by the columns of $X_{-v}$. Equation \eqref{eqGeneralP} in Proposition \ref{propbias} shows the parallels between the two approaches, although they were derived from different perspectives. In particular, when $a=\delta$ our operations on the sample space, designed to minimize a feasible upper bound on the mean squared error, recover the ridge regression residual vector for $z_v$, rather than the  nodewise lasso residuals from an $\ell_1$-constrained linear regression of $x_v$ on $X_{-v}$  recommended by \cite{ZZ2014} and \cite{vdG2014}. 

While there is no obvious direct analogue of our quantities $b_v$ or $(q_v^Tx_v)b_v/(\tau q_v^{T} q_v)^{1/2}$, \cite{vdG2014} show that their Studentized test statistic is the sum of a standard normally distributed random variable and the quantity $\widetilde{\Delta}_v$, satisfying
\[
\text{pr}\biggl(|\widetilde{\Delta}_v| \geq 8 n \biggl(\frac{ s\lambda_v \lambda}{\tau \phi_0^2\|x_v - X_{-v}\widehat{\gamma}_v\|_2}\biggr)\biggr) \leq 2 \exp(-t^2).
\]
Here, $\phi_0^2$ is the so-called compatibility constant of the lasso, $\widehat{\gamma}_v$ is the nodewise lasso estimator of the  coefficient vector from a regression of $x_v$ on $X_{-v}$ with regularization parameter $\lambda_v$, and $\lambda$ (chosen as a function of $t>0$) is the lasso tuning parameter associated with the regression of $Y$ on $X$. They also show that their debiased lasso estimator satisfies $\sqrt{n}(\widehat{\beta}^\text{d}-\beta)=W+\Delta$ where $W$ is a normal random variable of zero mean and $\Delta$ satisfies $\|\Delta\|_\infty=o_\text{pr}(1)$. This theoretical guarantee is established under an assumption of row-sparsity of the inverse Fisher information matrix. Our result of Proposition \ref{propBiasb} shows that $b_v = O(s/n)$ without the extra sparsity condition (beyond sparsity of $\beta$) and in a Gaussian setting induces sparsity on the relevant row and column of the Fisher information matrix, as in \cite{CR1987}. Thus our approach seeks to achieve by construction what is implicitly assumed in \cite{vdG2014} and \cite{ZZ2014}.  Inducement of  sparsity  cannot be accomplished simultaneously for all parameters, but is achievable by considering each coefficient in turn as the interest parameter.

Another important advantage of our approach is that it does not  require rescaling the columns of $X$ to unit length, so the physical interpretation of the  estimate and confidence intervals is maintained. Neither the lasso estimator nor its debiased counterpart is invariant to componentwise rescaling, and  simulations in \S \ref{secNearlyReal}  indicate that the approach is rather conservative after transformation to the original scale.

 \cite{JM2014} also consider a version of the debiased lasso estimator for the linear model. Their estimator is of a similar form to that of \cite{vdG2014} and \cite{ZZ2014}, namely
\[
\widehat{\beta}^{\text{JM}}=\widehat{\beta}_{\text{lasso}} + n^{-1}MX^T(Y-X\widehat{\beta}_{\text{lasso}}),
\]
where, in a similar spirit to our proposal, $M$ is chosen to minimize the variance of the resulting estimator subject to a constraint on the bias. Specifically, $M=(m_1,\ldots,m_p)^T$, where $m_v$ solves
\[
m_v = \argmin_m m^{T} (n^{-1}\textstyle{\sum_{i}}X_{i}X_{i}^T )m \quad \text{s.t. } \|(n^{-1}\textstyle{\sum_{i}}X_{i}X_{i}^T )m - e_v \|_{\infty} \leq \mu,
\]
with $e_v$ the zero vector with a 1 in the $v$th position. Thus $M$ can be viewed as a regularized inverse of $n^{-1}\textstyle{\sum_{i}}X_{i}X_{i}^T$. A key difficulty with this proposal is that the tuning parameter $\mu$ needs to decay sufficiently quickly in order that the bias of the estimator (the analogue of our $b_v$) may be controlled. \cite{JM2014} state in their Algorithm 1 that if any of the constraints is violated, then $M$ should be set to the identity matrix. This seems problematic since the constraint is used to derive the theoretical guarantees. We found that it was often violated in examples we tried.

\section{Confidence sets of models}\label{secConfModels}

\subsection{Background}\label{secConfBkgrd}

For sparse high-dimensional regression problems, \cite{CB2017} emphasised that several or even many low-dimensional models are likely to fit the data indistinguishably well. They argued that while an arbitrary choice between them would be reasonable for prediction, for subject-matter understanding, it is more appropriate to report a confidence set of models. Our suggested use for the estimates and confidence intervals obtained as described in  \S \ref{secCS} is as an adjunct to such confidence sets, the two procedures being performed in isolation, thereby avoiding issues of post-selection inference.

The first phase in the construction of confidence sets of models is to identify a set of retained explanatory variables, $\widehat{\mathcal{S}}$, where $|\widehat{\mathcal{S}}|$ is reasonably large, but considerably less than $n$. The model that includes all variables in $\widehat{\mathcal{S}}$  is called the encompassing model. A confidence set of models, $\mathcal{M}$, is obtained by comparing the fit of the encompassing model to the fit of any model using $s^{\#}$ or fewer variables from $\widehat{\mathcal{S}}$, and retaining any such models that are statistically indistinguishable from the encompassing one. The choice of $s^{\#}$ is arbitrary and based on an assumption of sparsity.

The confidence intervals in \S \ref{secCS} can be used to refine  this large confidence set, $\mathcal{M}$, to  a smaller set $\mathcal{R}\subset \mathcal{M}$,  by removing from $\mathcal{M}$ any models which have  coefficient estimates that are not included in the relevant confidence interval \eqref{eqCSPsi} (with $\tau$ replaced by $V_n$). Specifically, if any of the estimated coefficients from fitting a model in $\mathcal{M}$ exceed such confidence limits, this casts doubt on the plausibility of that model. If none of the models in $\mathcal{M}$ is consistent with the collection of confidence intervals, then this suggests that $|\widehat{\mathcal{S}}|$ is too small.

 We assume that $\mathcal{M}$ is exactly calibrated, i.e. that $\text{pr}(\mathcal{S} \in \mathcal{M})=1-\vartheta$. Asymptotic calibration of $\mathcal{M}$ is established under certain conditions in \citet{LewisBattey2021}. As discussed in \S \ref{secTheory}, while  the individual confidence intervals are all calibrated at their nominal levels  for sufficiently large $n$, small-sample  bias affects inference on some parameters more than others. The following calculations indicate the implications of this bias on the small-sample coverage properties of $\mathcal{R}$. 

Let $q = q(u)$ be the proportion of confidence intervals whose coverage probability is below some acceptable threshold, $1-u$ say. Since poor calibration is unrelated to whether a variable is in $\mathcal{S}$, the indices of signal variables can be treated as a simple random sample of size $s$ from $\{1,\ldots,p\}$. Thus, the number of poorly calibrated intervals among these $s$ is binomial of index $s$ and parameter $q$, and an approximate lower bound on the probability that no signal variable is excluded from $\mathcal{M}$ is $((1-q)(1-u))^s$.

The analysis of artificially generated responses and real covariate data in \S \ref{secNearlyReal} have $q = 23/4088 = 0.0056$ intervals with coverage smaller than $1-u=0.95$ when the nominal level of our intervals is $0.01$, $q = 118/4088 = 0.0289$  with coverage smaller than $1-u=0.98$ at nominal level $0.005$ and $q = 20/4088 = 0.0049$ with coverage smaller than $1-u=0.99$ at nominal level $0.001$ (see Table \ref{tableBuhlmann}). Thus, for $s=5$, and $\text{pr}(\mathcal{S} \in \mathcal{M})=0.99$, the coverage probability of $\mathcal{R}$ using the $0.001$-level intervals is approximately $((1-q)(1-u))^5(1-\vartheta)=0.9186$.

The above arguments would not work for the debiased lasso, as simulations in \S \ref{secFactorial} show that, while the coverage probabilities are high on average, the debiased lasso intervals are more systematically miscalibrated for signal variables.

Whether the refined set $\mathcal{R}$ is more useful than a version of $\mathcal{M}$ whose coverage matches $\mathcal{R}$ depends on $q$ and its associated value of $u$. This can be assessed through an analysis similar to that in \S \ref{secNearlyReal}.

\subsection{Illustration}\label{secIllustration}

\cite{BuhlmannAR} describe an investigation to study the association of the response variable, vitamin $B_2$ production, with the logarithmic expression levels of $4088$ genes. In the sample of size $n=71$ the correlation among columns of $X$ ranges from  $-0.926$ to $0.991$, with a mean correlation $0.029$. See  \cite{BuhlmannAR} for original references.

The first analysis is the estimation of each of the regression coefficients and their confidence limits at nominal level $\alpha=0.001$ using the approximate orthogonalization approach of \S \ref{secCS}. A separate analysis was carried out to choose the encompassing model $\widehat{\mathcal{S}}$ and a confidence set of models $\mathcal{M}$. The variables in $\widehat{\mathcal{S}}$ were identified through a series of model fits based on partially balanced incomplete block arrangements \citep{Yates1936, CB2017}. Then each possible low-dimensional sub-model of $\widehat{\mathcal{S}}$  was tested against the  encompassing model using an $F$-test at level $\vartheta=0.01$. Models not rejected by the $F$-test comprise the confidence set $\mathcal{M}$.

The confidence intervals constructed as described in \S \ref{secCS} for the 22 variables in $\widehat{\mathcal{S}}$ are reported in Table \ref{tableS} for $\alpha=0.05$ and $\alpha=0.001$. These have been ordered according to the proportion of models from $\mathcal{M}$ to which they belong. For comparison, the lasso and the elastic net, fitted to all 71 observations and tuned to retain 22 variables, find nine and fourteen of the same variables. These are indicated in the first column of Table \ref{tableS}. Further explanation of $\widehat{\mathcal{S}}$ and $\mathcal{M}$ is provided in Appendix \ref{secDetailsConfSetsModels}, including details of error variance estimation and how asymptotic calibration of $\mathcal{M}$ is ensured.

There are 34555 models in $\mathcal{M}$, but only 773 of these give estimates of the regression coefficients that are compatible with our nominal 99.9\% confidence intervals for the individual parameters in Table \ref{tableS}. Let $\mathcal{R}$ denote this set of 773 models. Since the decision to report a model in $\mathcal{R}$ depends on the simultaneous correctness of the associated confidence intervals, a Bonferroni-type correction should be applied as in \S \ref{secConfBkgrd}. In fact, to account for a degree of finite-sample miscalibration of our intervals due to the term $b_v$, we use the argument of \S \ref{secConfBkgrd} with $q$ and the corresponding value of $u$ estimated from Table \ref{tableBuhlmann}. Thus the effective coverage probability of $\mathcal{R}$ assuming, optimistically, that the coverage probability for $\mathcal{M}$ is exactly calibrated at its nominal level, is approximately $0.9186$. For comparison, the size of the model confidence set based only on a likelihood ratio test at nominal level $1-0.9186=0.0814$ is $15084$, illustrating the advantage of our intervals as an adjunct to confidence sets of models. Any choice between the 773 models in $\mathcal{R}$ would require either additional data or subject matter expertise.  Table \ref{tableM2} reports central regions of the confidence ``distribution'' of models \citep[in the sense of][]{Fisher1930, Cox1958} obtained as a nested sequence of confidence sets at different levels. The full confidence set of models is reported in the supplementary material. 
\begin{table}
	{\centering		
		\begin{tabular}{c|cccccc}
			variable & proportion & \multirow{2}{2pt}{$\widetilde{\beta}_v$} & lower limit & upper limit & lower limit & upper limit \\ 
			index $v$ & of models in $\mathcal{M}$  & & (0.05) & (0.05) & (0.001) & (0.001) \\ 
			\hline 			 
			2138 & 0.218 & -0.062 & -0.382 & \hspace{4pt}0.259& -0.599 &\hspace{4pt}0.476\\ 
			2564$^{L,E}$ & 0.218 & -1.481 & -1.801 & -1.160 & -2.018& -0.943\\ 
			1516$^{L,E}$ & 0.218 & \hspace{4pt}0.343 & \hspace{4pt}0.022 & \hspace{4pt}0.663  & -0.195& \hspace{4pt}0.880\\
			1503$^{L,E}$ & 0.217 & -0.325 & -0.646 & \hspace{6pt}-0.0050 & -0.863&\hspace{4pt}0.212\\ 
			1639$^{L,E}$ & 0.217 & -0.406 & -0.726 & -0.086 &-0.944&\hspace{4pt}0.132\\ 
			1603 & 0.217 & -1.048 & -1.368 & -0.728 &-1.586& -0.510\\ 
			4008$^E$ & 0.217 & -0.366 & -0.686 & -0.046 &-0.903&\hspace{4pt}0.172\\ 
			4002$^{L,E}$ & 0.216 & -0.505 & -0.825 & -0.185 &-1.043&\hspace{4pt}0.033\\ 
			1069$^E$ & 0.216 & -0.398 & -0.718 & -0.078 &-0.936&\hspace{4pt}0.140\\ 
			1436$^E$ & 0.215 & -0.463 & -0.783 & -0.143 &-1.001& \hspace{4pt}0.075 \\ 
			3291 & 0.215 & -0.640 & -0.960 & -0.320 &-1.178&-0.102\\ 
			978  & 0.214 & -0.259 & -0.580 & \hspace{4pt}0.061 &-0.797&\hspace{4pt}0.278\\ 
			3514$^{L,E}$ & 0.214 & \hspace{4pt}1.373 & \hspace{4pt}1.053 & \hspace{4pt}1.694 &\hspace{4pt}0.836& \hspace{4pt}1.911 \\
			1297$^{L,E}$ & 0.213 & \hspace{4pt}0.219 & -0.102 & \hspace{4pt}0.539 &-0.319& \hspace{4pt}0.756\\ 
			1285 & 0.213 & \hspace{4pt}0.172 & -0.148 & \hspace{4pt}0.493 &-0.366& \hspace{4pt}0.710\\ 
			3808$^E$ & 0.213 & \hspace{4pt}0.677 & \hspace{4pt}0.356 & \hspace{4pt}0.997 &\hspace{4pt}0.139& \hspace{4pt}1.214\\ 
			1423 & 0.212 & \hspace{4pt}0.043 & -0.277 & \hspace{4pt}0.363 &-0.495& \hspace{4pt}0.581\\ 
			1278$^{L,E}$ & 0.212 & \hspace{4pt}0.147 & -0.173 & \hspace{4pt}0.467 &-0.391& \hspace{4pt}0.685\\ 
			403  & 0.211 & \hspace{4pt}0.902 & \hspace{4pt}0.582 & \hspace{4pt}1.44 &\hspace{4pt}0.365&\hspace{4pt}1.323\\ 
			1290 & 0.211 & \hspace{4pt}0.189 & -0.131 & \hspace{4pt}0.510 &-0.348&\hspace{4pt}0.727\\ 			
			1303$^E$ & 0.211 & \hspace{4pt}0.187 & -0.133 & \hspace{4pt}0.507 &-0.351&\hspace{4pt}0.725\\ 
			1312$^{L,E}$ & 0.209 & \hspace{4pt}0.490 & \hspace{4pt}0.169 & \hspace{4pt}0.810 &\hspace{4pt}0.048&\hspace{4pt}1.027\\ 
			\hline 
		\end{tabular}
		\caption{Indices for variables in $\widehat{\mathcal{S}}$, ordered by prevalence in $\mathcal{M}$, plus 95\% and 99.9\% confidence limits for the associated regression coefficients. Superscripts $L$ and $E$ indicate that the variable was also found by the lasso and the elastic net fitted to the full sample. \label{tableS}}
	}
\end{table}

{\centering	
	\begin{table}
		\vspace{-1cm}
		\begin{tabular}{ccccccccccccccccccccccc}
			&      \begin{sideways}2138\end{sideways} & \begin{sideways}2564\end{sideways} & \begin{sideways}1516\end{sideways} & \begin{sideways}1503\end{sideways} & \begin{sideways}1639\end{sideways} & \begin{sideways}1603\end{sideways} & \begin{sideways}4008\end{sideways} & \begin{sideways}4002\end{sideways} & \begin{sideways}1069\end{sideways} & \begin{sideways}1436\end{sideways} & \begin{sideways}3291\end{sideways} & \begin{sideways}978\end{sideways} & \begin{sideways}3514\end{sideways} & \begin{sideways}1297\end{sideways} & \begin{sideways}1285\end{sideways} & \begin{sideways}3808\end{sideways} & \begin{sideways}1423\end{sideways} & \begin{sideways}1278\end{sideways} & \begin{sideways}403\end{sideways} & \begin{sideways}1290\end{sideways} & \begin{sideways}1303\end{sideways} & \begin{sideways}1312\end{sideways} \\
			
			\hline 
1&-&	-&	-&	-&	-&	-&	-&	-&	-&	-&	-&	*&	-&	*&	-&	-&	-&	-&	-&	*&	-&	-\\
2&-&	-&	-&	-&	-&	-&	-&	-&	*&	-&	-&	-&	-&	-&	*&	-&	-&	*&	-&	-&	-&	-\\
3&-&	-&	-&	-&	-&	-&	-&	-&	*&	-&	-&	-&	-&	*&	-&	-&	-&	*&	-&	-&	-&	-\\
4&-&	-&	-&	-&	-&	-&	-&	-&	*&	-&	-&	-&	-&	-&	-&	-&	*&	*&	-&	-&	-&	-\\
5&-&	-&	-&	-&	-&	-&	-&	-&	*&	-&	-&	-&	-&	-&	*&	-&	*&	-&	-&	-&	-&	-\\
6&-&	-&	-&	-&	-&	-&	-&	-&	*&	-&	-&	*&	-&	-&	*&	-&	-&	*&	-&	-&	-&	-\\
7&-&	-&	-&	-&	-&	-&	-&	-&	-&	-&	-&	*&	-&	*&	*&	-&	-&	*&	-&	-&	-&	-\\
8&-&	-&	-&	-&	-&	-&	-&	-&	-&	-&	-&	*&	-&	*&	*&	-&	*&	-&	-&	-&	-&	-\\
9&-&	-&	-&	*&	*&	-&	-&	-&	-&	-&	-&	*&	-&	-&	-&	*&	-&	-&	-&	-&	-&	-\\
\rowcolor{aliceblue}
10&-&	-&	-&	-&	-&	-&	-&	-&	*&	-&	-&	-&	-&	*&	-&	-&	-&	-&	-&	-&	*&	-\\
\rowcolor{aliceblue}
11&-&	-&	-&	*&	-&	-&	-&	-&	*&	-&	-&	-&	*&	-&	-&	-&	-&	-&	-&	-&	-&	-\\
\rowcolor{aliceblue}
12&-&	-&	-&	*&	-&	-&	-&	-&	-&	*&	-&	-&	*&	-&	-&	-&	-&	-&	-&	-&	-&	-\\
\rowcolor{aliceblue}
13&-&	-&	-&	*&	-&	*&	*&	-&	-&	-&	-&	-&	-&	-&	-&	-&	-&	-&	-&	-&	-&	-\\
\rowcolor{interpolateblue}
14&-&	-&	-&	-&	-&	-&	-&	-&	*&	-&	-&	-&	-&	*&	-&	-&	-&	-&	-&	*&	-&	-\\
\rowcolor{interpolateblue}
15&-&	-&	-&	*&	*&	-&	-&	-&	-&	-&	-&	-&	-&	-&	-&	*&	-&	-&	-&	-&	-&	-\\
\rowcolor{beaublue}
16&-&	-&	-&	-&	-&	-&	-&	-&	*&	-&	-&	-&	-&	-&	-&	-&	*&	-&	-&	-&	-&	*\\
\rowcolor{beaublue}
17&-&	-&	-&	-&	-&	-&	-&	-&	*&	-&	-&	-&	-&	*&	*&	-&	-&	*&	-&	-&	-&	-\\
\rowcolor{beaublue}
18&-&	-&	*&	*&	-&	-&	-&	-&	-&	-&	-&	-&	-&	-&	-&	*&	-&	-&	-&	-&	-&	*\\
\rowcolor{beaublue}
19&-&	-&	-&	-&	-&	-&	-&	-&	*&	-&	-&	*&	-&	*&	*&	*&	-&	-&	-&	-&	-&	-\\
\rowcolor{interpolateblue}
20&-&	-&	*&	-&	-&	*&	-&	*&	-&	-&	-&	-&	-&	-&	-&	-&	-&	-&	-&	-&	-&	-\\
\rowcolor{interpolateblue}
21&-&	-&	-&	-&	-&	-&	-&	-&	*&	-&	-&	-&	-&	*&	*&	*&	-&	-&	-&	-&	-&	-\\
\rowcolor{aliceblue}
22&-&	-&	-&	-&	-&	-&	-&	-&	*&	-&	-&	-&	-&	*&	-&	-&	-&	-&	-&	*&	*&	-\\
\rowcolor{aliceblue}
23&-&	-&	-&	-&	-&	-&	-&	-&	*&	-&	-&	-&	-&	*&	-&	-&	*&	-&	-&	-&	-&	*\\
\rowcolor{aliceblue}
24&-&	-&	-&	*&	*&	-&	-&	-&	*&	*&	-&	-&	-&	-&	-&	*&	-&	-&	-&	-&	-&	-\\
\rowcolor{aliceblue}
25&-&	-&	*&	*&	-&	-&	-&	-&	-&	*&	-&	-&	-&	-&	-&	*&	-&	-&	-&	-&	-&	*\\
26&-&	-&	-&	-&	-&	-&	-&	-&	*&	-&	-&	-&	-&	*&	-&	-&	*&	*&	-&	-&	-&	-\\
27&-&	-&	-&	-&	*&	*&	-&	*&	-&	*&	-&	-&	-&	-&	-&	-&	-&	-&	-&	-&	-&	-\\
28&-&	-&	-&	-&	-&	-&	-&	-&	*&	-&	-&	*&	-&	*&	*&	-&	-&	*&	-&	-&	-&	-\\
29&-&	-&	-&	*&	-&	-&	-&	-&	*&	-&	-&	*&	-&	-&	-&	-&	-&	*&	-&	-&	-&	*\\
30&-&	-&	-&	*&	-&	-&	-&	-&	*&	-&	-&	*&	-&	-&	*&	-&	-&	-&	-&	-&	-&	*\\
31&-&	-&	-&	*&	-&	-&	-&	-&	*&	-&	-&	*&	-&	-&	-&	-&	-&	-&	-&	*&	-&	*\\
32&-&	-&	-&	-&	-&	-&	-&	-&	*&	-&	-&	-&	-&	*&	*&	-&	*&	-&	-&	-&	-&	*\\
33&-&	-&	-&	*&	*&	-&	-&	-&	*&	-&	*&	-&	-&	-&	-&	*&	-&	-&	-&	-&	-&	-\\
34&-&	-&	*&	*&	-&	-&	-&	-&	-&	*&	-&	-&	-&	*&	-&	*&	-&	-&	-&	-&	-&	-\\
			\hline
		\end{tabular}
		\medskip
		\captionsetup{width=1.1\linewidth}
		\caption{Central region of a confidence distribution of models, obtained from successively smaller values of $\alpha$.  Shading corresponds to (from dark to light): $\alpha\in\{0.5,0.4,0.3,0.2\}$. The full confidence set of models $\mathcal{R}$ with coverage probability $0.9186$ is reported as supplementary material.
			\label{tableM2}}
	\end{table}
	
}

\section{Simulations}\label{secSims}

\subsection{Using covariate data from \citet{BuhlmannAR}}\label{secNearlyReal}

 In this section we use the covariate data from Section \ref{secIllustration} to study the coverage and bias in a moderately realistic example, and compare the results to the debiased lasso. Simulations of this nature are a helpful adjunct to the analysis above, to assess the security of the results. 

\begin{figure}[h]
	\begin{center}
		\includegraphics[trim=0.4in 3in 0.8in 3.3in, clip,width=0.49\linewidth, height=0.4\linewidth]{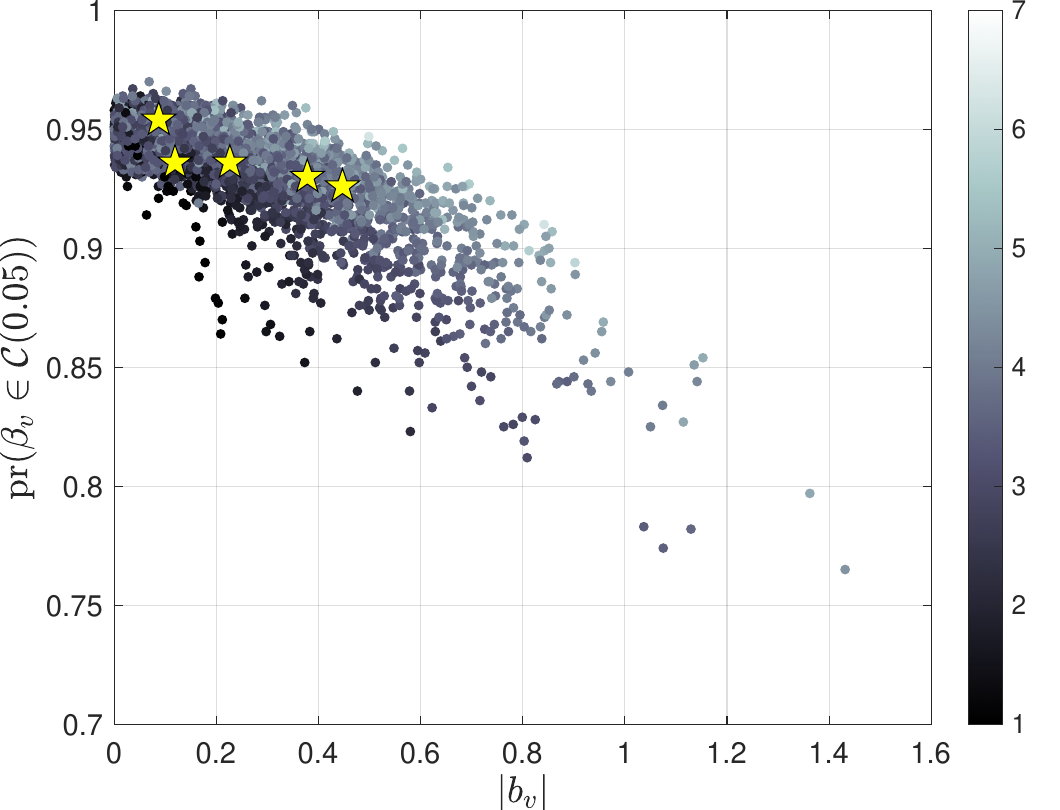} 		\includegraphics[trim=0.4in 3in 0.8in 3.3in, clip,width=0.49\linewidth, height=0.4\linewidth]{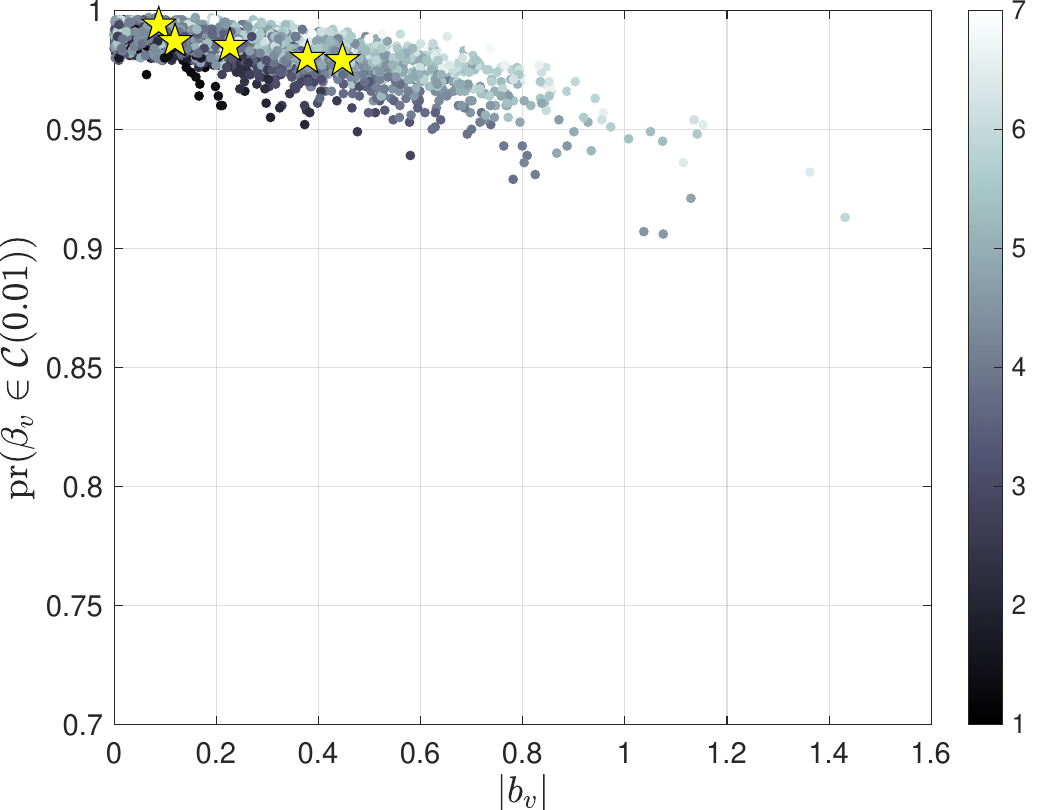}
	\end{center}
	\vspace{-0.4cm}
	\caption{Real covariate data with artificially generated outcomes: simulated coverage probabilities plotted against the absolute bias $|b_v|$ of the estimator with the the mean length of the confidence interval over Monte Carlo replications indicated by the shading. Nominal levels are $0.05$ (left) and $0.01$ (right) \label{figBuhlmann}}
\end{figure}

We drew indices for five signal variables at random from the index set $\{1,\ldots,p\}$, giving $\mathcal{S}=\{313, 689, 724, 1747, 2470\}$, fixed across 1000 ensuing Monte Carlo replications. In each, an artificial outcome was generated as ${Y=X\beta + \varepsilon}$, where $\varepsilon$ is a vector of independent standard normally distributed random variables. The signal strength was 1 for all signal variables. The assumption of known error variance used here was relaxed in \S \ref{secIllustration}. Confidence intervals at the $\alpha\in\{0.05,0.01,0.005,0.001\}$ nominal levels for  each of the $4088$ coefficients were constructed  as described in \S \ref{secCS}, and the coverage probabilities and mean lengths for each coefficient were estimated as simulation averages. Modal and median simulated coverage probabilities over the $4088$ variables and the median of the simulated mean lengths are reported in Table \ref{tableBuhlmann}.  Detailed summary information is depicted in Figure \ref{figBuhlmann}, where the coverage probabilities are plotted against $|b_v|$ and mean length. Yellow stars indicate signal variables. As expected, intervals with higher coverage (\ref{figBuhlmann}, right) are on average longer, and the actual coverage decreases as the bias increases, although relatively slowly.

\bigskip

\begin{table}[h]
	\begin{center}
		\begin{tabular}{c|ccccccc}
			nominal   & modal      & median  & median & \multicolumn{4}{c}{$1-u$}  \\ 
			\cline{5-8}
			level &  coverage & coverage & length & $0.9$ & $0.95$ &  $0.98$ & $0.99$  \\
			\hline
			0.05 & 0.951 & 0.942 & 3.32 & 0.939 & 0.228 & 0 & 0 \\
			0.01 & 0.989 & 0.987 & 4.37 & 1 & 0.994 & 0.839 & 0.288 \\
			0.005 & 0.989 & 0.987 & 4.77 & 1 & 0.999 & 0.971 & 0.798 \\
			0.001 & 0.995 & 0.993 & 5.59 & 1 & 1 & 1 & 0.995 \\
			\hline
		\end{tabular}
		\caption{Summary statistics over the 4088 coefficients of the simulated coverage probabilities and simulated mean lengths for each confidence interval. The last four columns are the values of $1-q$ used in the calculation of \S \ref{secConfBkgrd}. \label{tableBuhlmann}}
	\end{center}
\end{table}

\begin{figure}[h]
	\begin{center}
		\includegraphics[trim=0.4in 3in 0.8in 3.3in, clip,width=0.49\linewidth, height=0.4\linewidth]{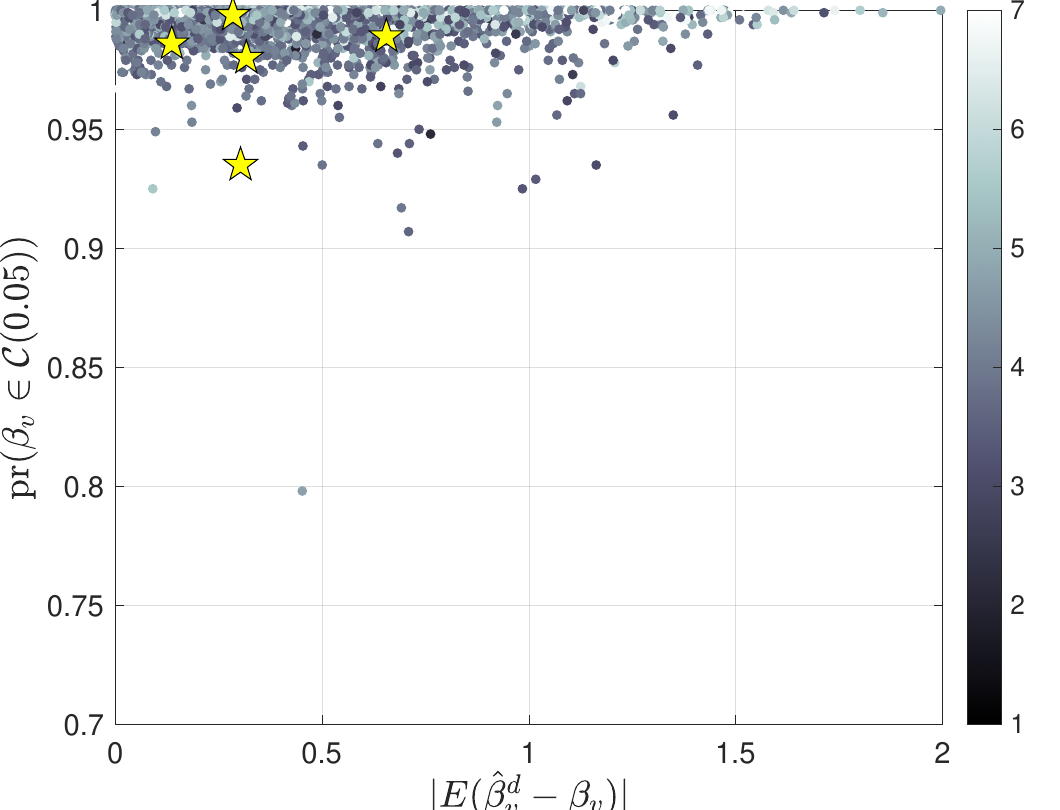} 		\includegraphics[trim=0.4in 3in 0.8in 3.3in, clip,width=0.49\linewidth, height=0.4\linewidth]{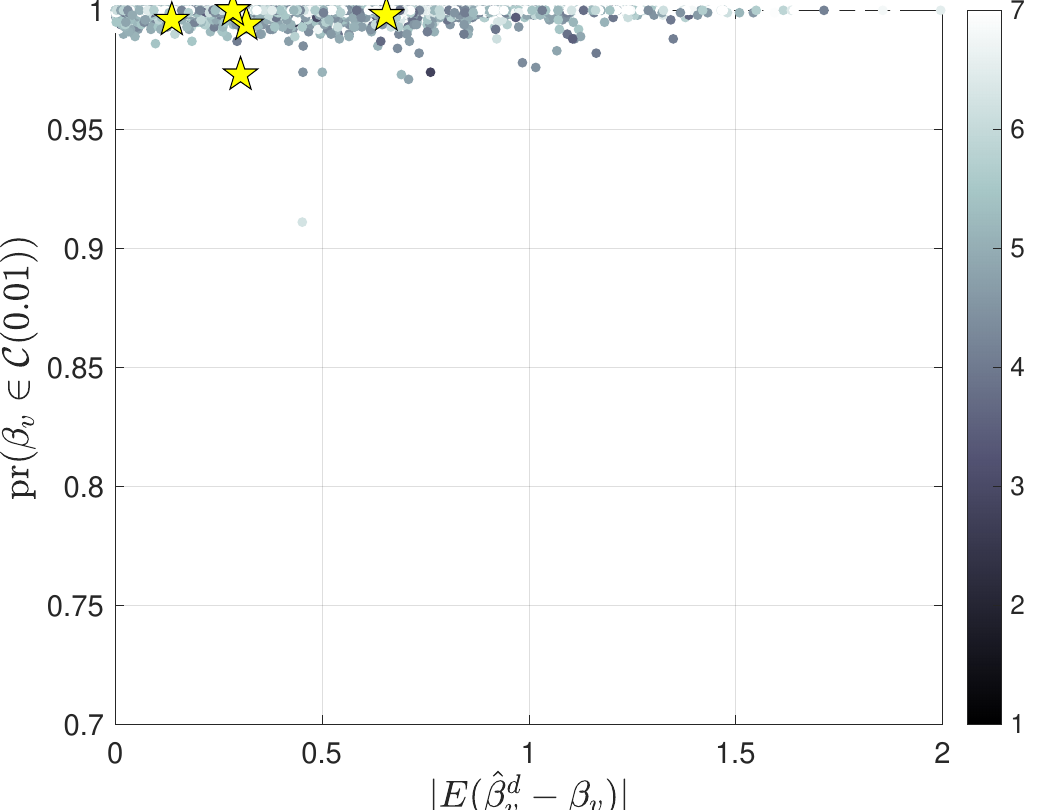}
	\end{center}
	\vspace{-0.4cm}
	\caption{Analogue of Figure \ref{figBuhlmann} for debiased lasso-based confidence intervals. \label{figBuhlmannDB}}
\end{figure}

For comparison, we also recorded the coverage properties and lengths for the debiased lasso estimator $\widehat{\beta}^{\text{d}}$ \citep{vdG2014} which, for the linear model, is essentially the same estimator as that of \cite{ZZ2014}. Since lasso-based procedures require the columns of $X$ to be on the same scale, we standardized them to unit length before constructing the debiased lasso in the usual way. In order to compare them to our approach, we then converted the conclusions back to the scale of interest by dividing the value of each entry of $\widehat{\beta}^{\text{d}}$ by the length of the corresponding column of the original $X$ matrix. Estimates of standard errors were adjusted by dividing them by the square root of this length. While more appropriate than a debiased lasso analysis on the original scale, the rescaling operation should not be expected to be adequate, as neither Wald-based inference nor the debiased lasso is invariant to rescaling. Summary statistics are reported in Table \ref{tableBuhlmannDB} and Figure \ref{figBuhlmannDB}. This demonstrates a degree of over-coverage, with the lengths of each interval tending to be larger than is necessary to achieve the nominal level (by comparison with Table \ref{tableBuhlmann}). Section \ref{secFactorial} gives a further comparison to the debiased lasso when the explanatory variables are all of similar magnitudes, so that no rescaling is required. The empirical coverage probabilities in that idealized setting are close to nominal, although the procedure appears to be systematically miscalibrated for signal variables, which is problematic for the type of analysis discussed in \S \ref{secConfBkgrd}.

\bigskip

\begin{table}[h]
	\begin{center}
		\begin{tabular}{c|ccccccc}
			nominal   & modal      & median & median & \multicolumn{4}{c}{$1-u$}  \\ 
			\cline{5-8}
			level &  coverage & coverage & length & $0.9$ & $0.95$ &  $0.98$ & $0.99$   \\
			\hline
			0.05 & 1 & 0.998 & 4.29 & 1 & 0.996 & 0.964 & 0.889 \\
			0.01 & 1 & 1 & 5.64 & 1 & 1 & 0.998 & 0.992 \\
			\hline
		\end{tabular}
		\caption{Analogue of first two rows of Table \ref{tableBuhlmann} for debiased lasso-based confidence intervals. \label{tableBuhlmannDB}}	
	\end{center}
\end{table}

We close this comparison with a comment on computation. Construction of our confidence intervals for a single component $\beta_v$ only entails construction of $q_v$, which is essentially instantaneous because of its closed-form solution. The computational burden associated with construction of a single interval is therefore approximately independent of $p$. By contrast, specification of the debiased lasso estimator and its variance estimate for $\beta_v$ entails construction of the whole matrix $\widehat{\Theta}_\text{Lasso}$ from equation (8) of \cite{vdG2014}. This involves solving $p$ nodewise lasso regressions, whose tuning parameters are ideally chosen by cross validation as in the experiments above. Specifically, there are $p+1000$ tuning parameters chosen by cross-validation involved in the generation of Figure \ref{figBuhlmannDB}, compared to one tuning parameter fixed at $\delta=1$ across Monte Carlo replicates for Figure \ref{figBuhlmann}, since the performance of our approach was found to be stable with respect to $\delta$, provided that a value extremely close to zero was not used. The computation of the confidence set $\mathcal{M}$ of \S \ref{secConfModels} is computationally more intensive as this entails checking all sub-models of dimension $\leq s^\#$ of the encompassing model for their compatibility with the data. With $s^\#=5$, computation can be performed within a minute on a standard computer for $|\hat{\mathcal{S}}|=22$ but computation time increases rapidly with $|\hat{\mathcal{S}}|$, and to a lesser extent with $s^\#$.

\subsection{A factorial experiment}\label{secFactorial}

In this section  we assess how coverage properties and lengths of the proposed intervals are affected by $p$, $n$ and the correlation between columns of $X$. In particular, we seek to understand the small-sample behaviour of the procedure in various circumstances.

Our Monte Carlo experiment was designed as follows. An $X$ matrix was obtained by generating each row $x_{i}^{T}$ for $i=1,\ldots, n$ from a normal distribution of mean zero and covariance matrix $\rho 1_{p}1_{p}^{T} + (1-\rho)I$,  and the columns were each centered to have sample mean zero.   The matrix $X$ was fixed across subsequent Monte Carlo replications, as was a sparse vector $\beta$, consisting of ones in the last five entries and zeros elsewhere. The values of $n$, $p$ and $\rho$ were taken at all $2^3$ combinations of high and low levels as in Table \ref{tableFactorial}, the value of $n/p$ being the same in two of the four configurations at high and low values of $\rho$. In each replication, a vector $\varepsilon$ of independent standard normally distributed errors was generated anew and a vector of outcomes constructed as ${Y=X\beta + \varepsilon}$. Each of the first 1000 entries of $\beta$ was treated in turn as the interest parameter and the coverage properties and lengths, obtained as in \S \ref{secNearlyReal}, are reported in Table \ref{tableFactorial}. These  set ${a=\delta=1}$ in formula \eqref{eqExpressionq}.  Simulations (not reported here) for ${a=1}$ and ${\delta=n^{-1}}$, produced identical numbers to those in Table \ref{tableFactorial} up to three significant figures.
\begin{table}
	\begin{center}
		\begin{tabular}{ccc|cccccc}
					&     &     &	modal   & median & proportion with & median & {median} & {95th p.c.} \\
 			$\rho$  & $n$ & $p$ &  coverage & coverage & coverage $>0.9$ & length & {$\vartheta_{w}^2$} & {$\vartheta_{w}^2$ } \\
			\hline
			0.9& 70 & 2450     & 0.987 & 0.980 & 0.922 & 1.975  & 0.0065 & 0.056 \\
			
			0.9& 70 & 1225    & 0.990  & 0.980 & 0.924 & 2.000 & 0.0063 & 0.055 \\
			
			0.9& 35 & 2450    & 0.986  & 0.979 & 0.898 & 2.786 & 0.0133 & 0.118\\
			
			0.9& 35 & 1225   & 0.988  & 0.981 & 0.937 &  2.801 & 0.0133 & 0.118 \\
			
			0.1& 70 & 2450    & 0.989 & 0.870 & 0.454 &  0.663 & 0.0065 & 0.056 \\
			
			0.1&70 & 1225   & 0.987  & 0.906 & 0.508 & 0.671  & 0.0063 & 0.054 \\
			
			0.1&35 & 2450   & 0.991  & 0.836 & 0.415 &  0.936 & 0.0135 & 0.120 \\
			
			0.1& 35 & 1225  & 0.988 & 0.917  & 0.538 &  0.947 & 0.0133 & 0.120 \\
			\hline 
		\end{tabular}
		\caption{Summary statistics for simulated coverage probabilities and mean lengths of $\alpha=0.01$ confidence intervals for the first 1000 coefficients (null variables only). The last two columns are the median and 95th percentile of the $1000\times (p-1)$ values of $\vartheta_w^2$ for the first 1000 coefficients.  \label{tableFactorial}}
	\end{center}
\end{table}

Table \ref{tableContrasts} reports point estimates of the main effects of $\rho$, $n$ and $p$,  from logistic regression of the coverage probabilities, and linear regression of the length, using the values in Table \ref{tableFactorial}.  For the coverage probabilities, the multiplicative effects are reported, with estimates close to one indicating a negligible effect on the coverage. The modal coverage probabilities are essentially unaffected by $\rho$, $n$ and $p$, as would be expected, as for most of the coefficients the approximate orthogonalization is effective and the statistic used is close to the pivotal one. At these small values of $n$, an increase from $n=35$ to $n=70$ reduces the length of the intervals but appears to have a negligible effect on the median coverage probability, while increasing $p$ has no effect on the length but multiplies the odds associated with the median coverage probabilities by $0.624$. This is to be expected: in optimizing the composite non-orthogonality over more variables, the signal variables play a relatively smaller role in the optimization problem. By far the largest effect on the median coverage is due to $\rho$. The median length of the confidence intervals also increases with the correlation, reflecting expression \eqref{eqqPsi} for the variance of $\widetilde{\beta}_v$. See Table \ref{tableIncreasingN} for further analysis on the role of $n$.

\begin{figure}
	\vspace{-1.5cm}
	\begin{center}
		$\rho = 0.9$ \hspace{6.5cm} $\rho = 0.1$ \\
	\end{center}
	\begin{subfigure}{0.01\textwidth}
		\begin{sideways}$n$ high, $p$ high
		\end{sideways}
	\end{subfigure}
	\begin{subfigure}{0.49\textwidth}
		\centering
		\includegraphics[trim=0.9in 3in 0.85in 3.3in, clip,width=.88\linewidth, height=0.75\linewidth]{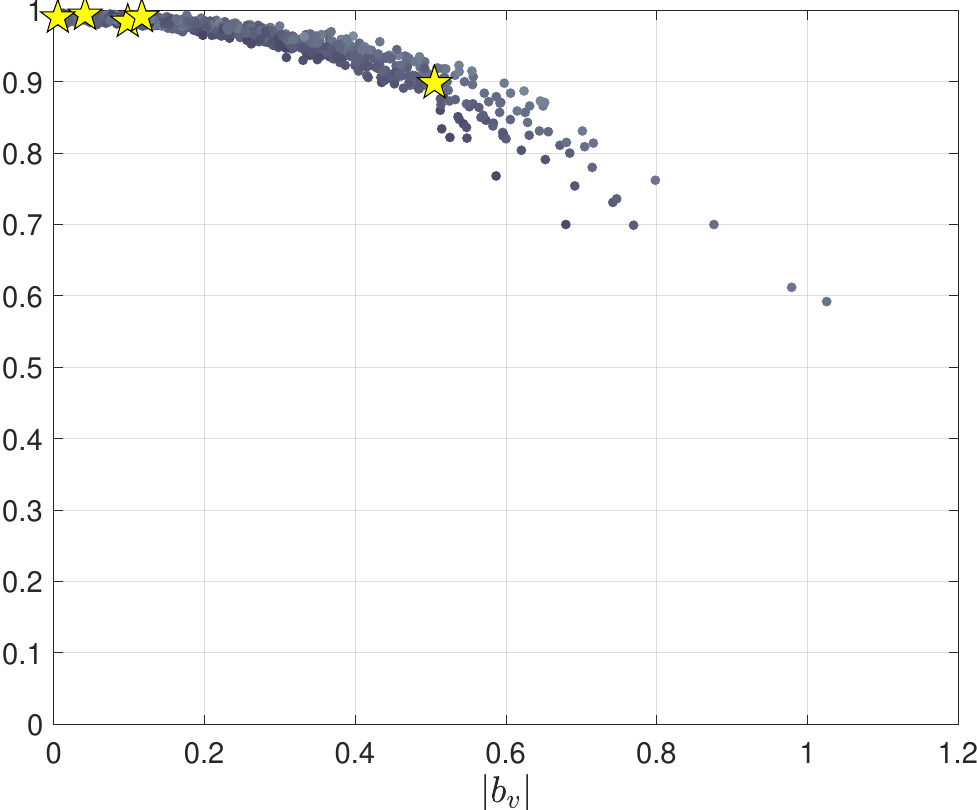}
	\end{subfigure}
	\begin{subfigure}{0.49\textwidth}
		\centering
		\includegraphics[trim=0.5in 3in 0.8in 3.3in, clip,width=\linewidth, height=0.75\linewidth ]{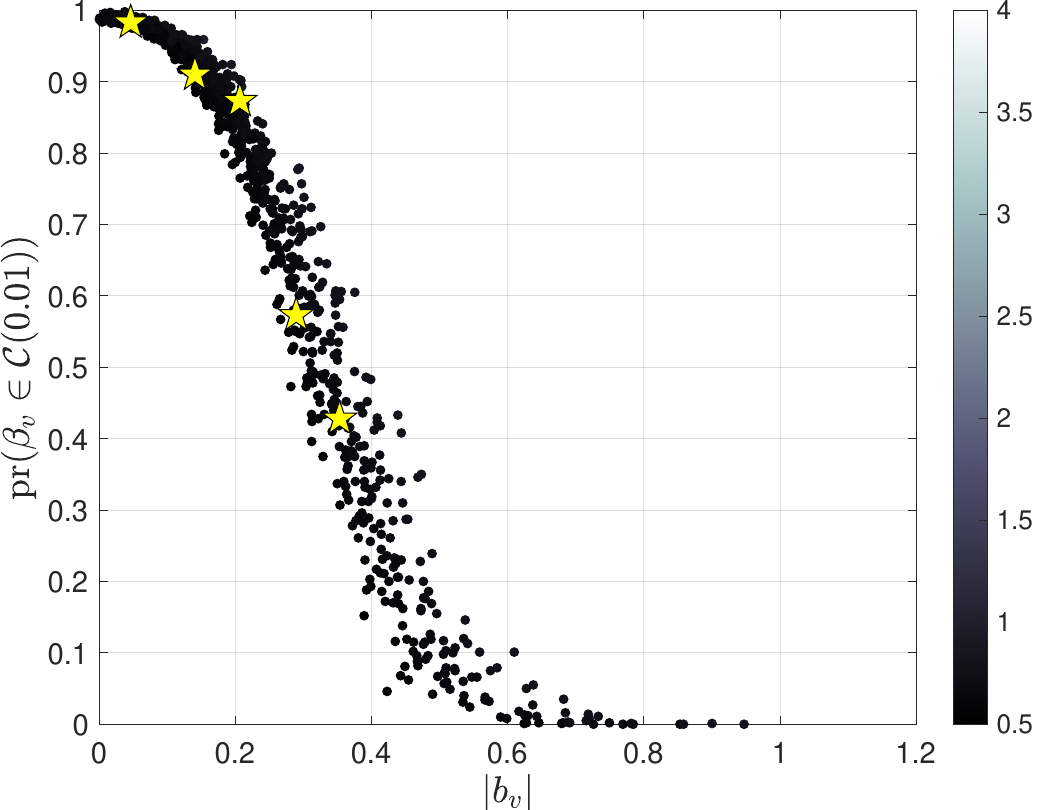} \\
	\end{subfigure}
	\begin{subfigure}{0.01\textwidth}
		\begin{sideways}$n$ high, $p$ low
		\end{sideways}
	\end{subfigure}
	\begin{subfigure}{0.49\textwidth}
		\centering
		\includegraphics[trim=0.9in 3in 0.85in 3.3in, clip,width=.88\linewidth, height=0.75\linewidth]{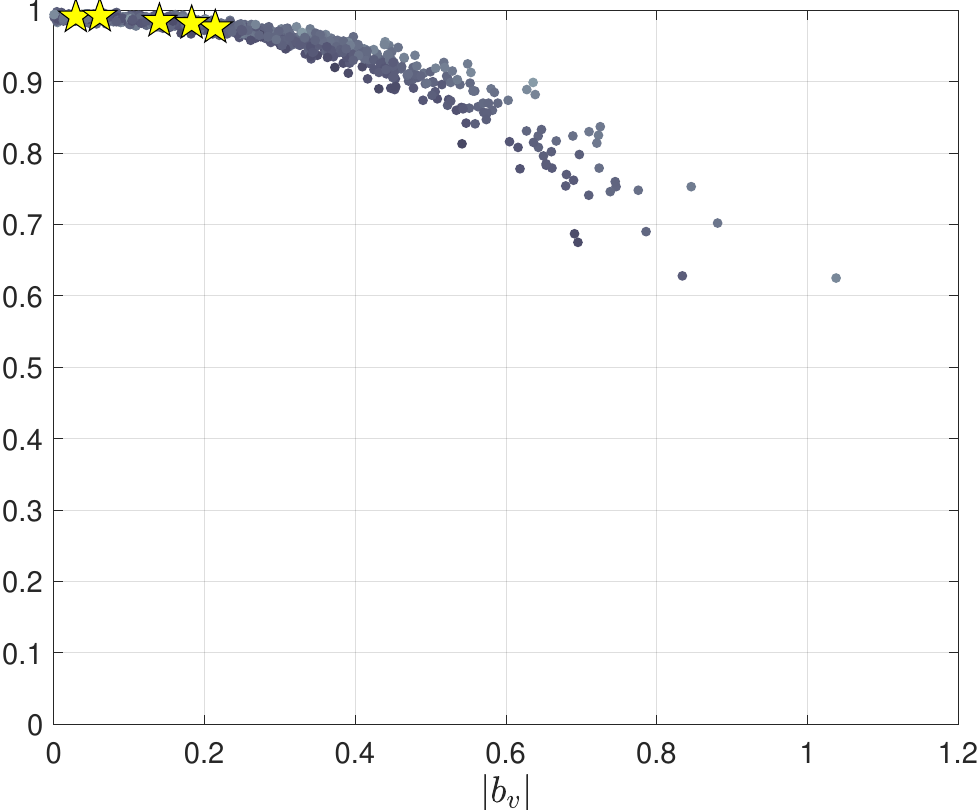}
	\end{subfigure}
	\begin{subfigure}{0.49\textwidth}
		\centering
		\includegraphics[trim=0.5in 3in 0.8in 3.3in, clip,width=\linewidth, height=0.75\linewidth]{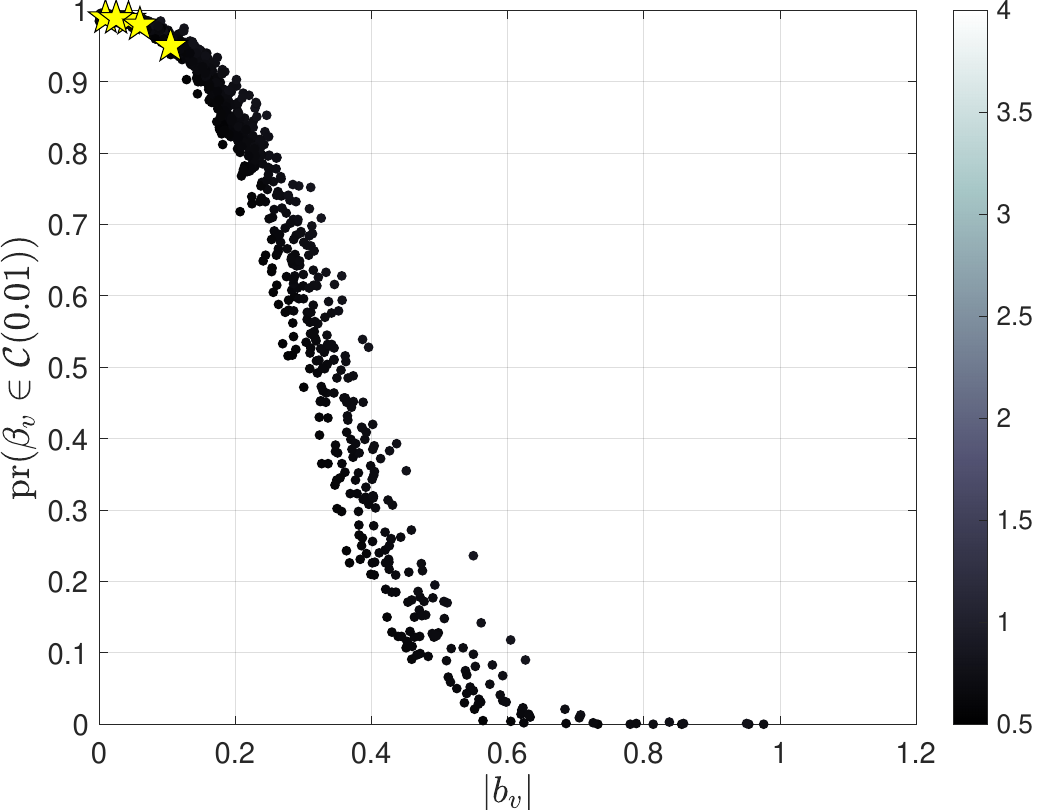} \\
	\end{subfigure}
	\begin{subfigure}{0.01\textwidth}
		\begin{sideways}$n$ low, $p$ high
		\end{sideways}
	\end{subfigure}
	\begin{subfigure}{0.49\textwidth}
		\centering
		\includegraphics[trim=0.9in 3in 0.85in 3.3in, clip,width=.88\linewidth, height=0.75\linewidth]{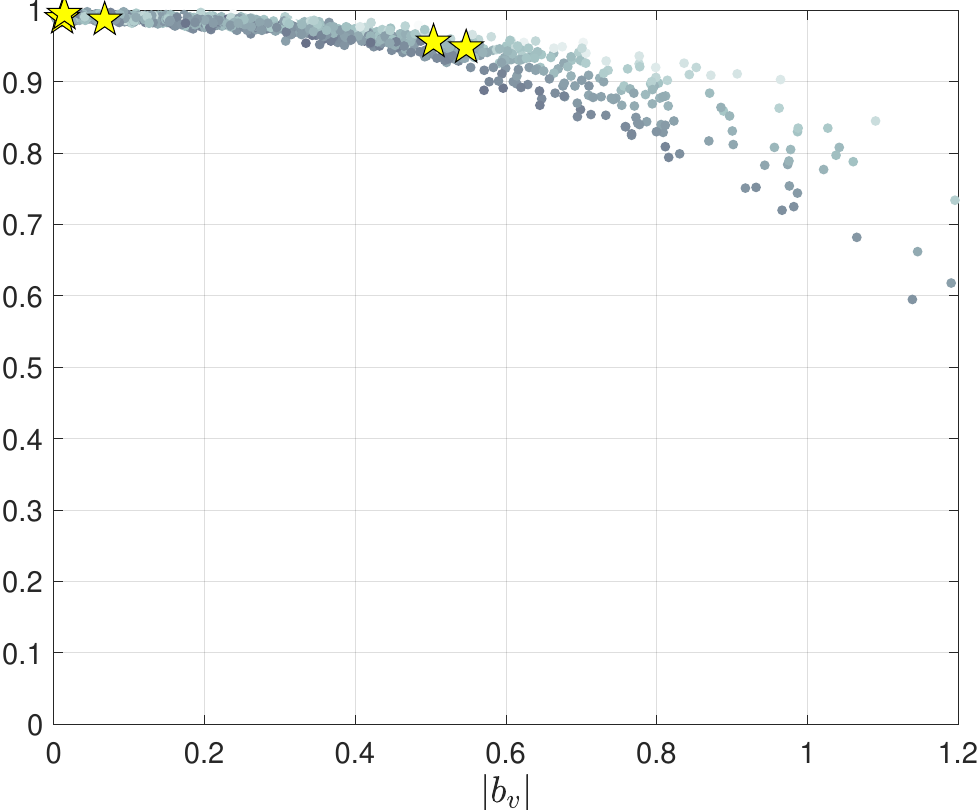}
	\end{subfigure}
	\begin{subfigure}{0.49\textwidth}
		\centering
		\includegraphics[trim=0.5in 3in 0.8in 3.3in, clip,width=\linewidth, height=0.75\linewidth]{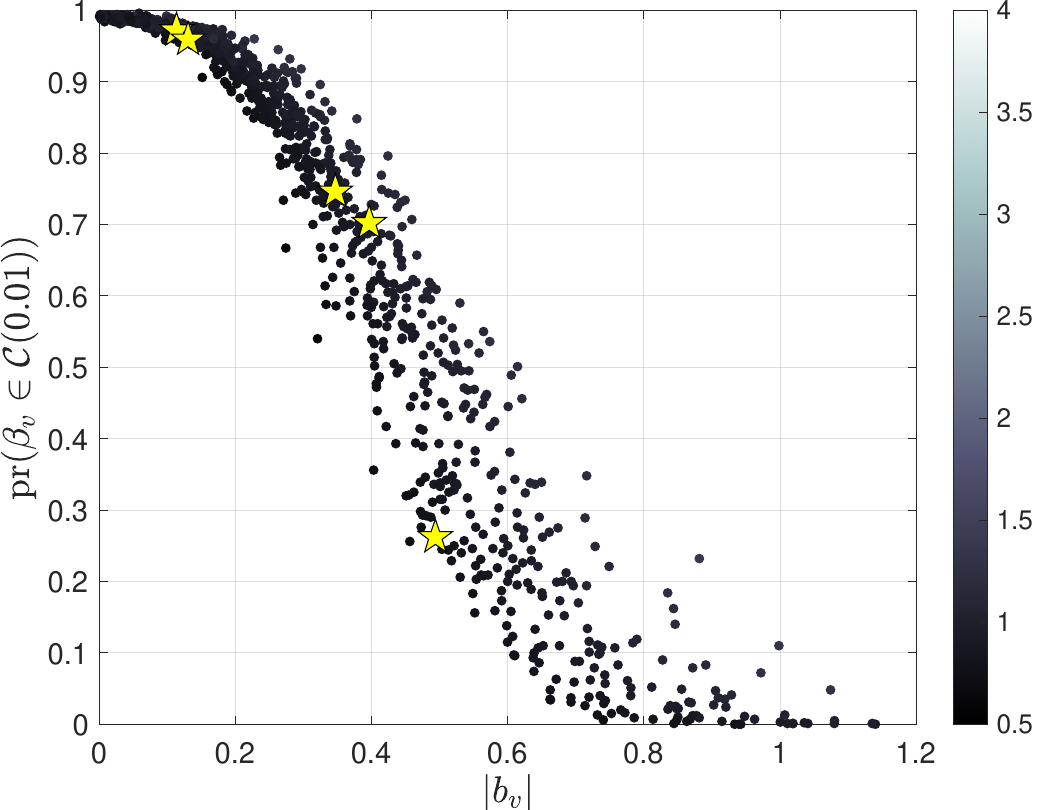} \\
	\end{subfigure}
	\begin{subfigure}{0.01\textwidth}
		\begin{sideways}$n$ low, $p$ low
		\end{sideways}
	\end{subfigure}
	\begin{subfigure}{0.49\textwidth}
		\centering
		\includegraphics[trim=0.9in 3in 0.85in 3.3in, clip,width=.88\linewidth, height=0.75\linewidth]{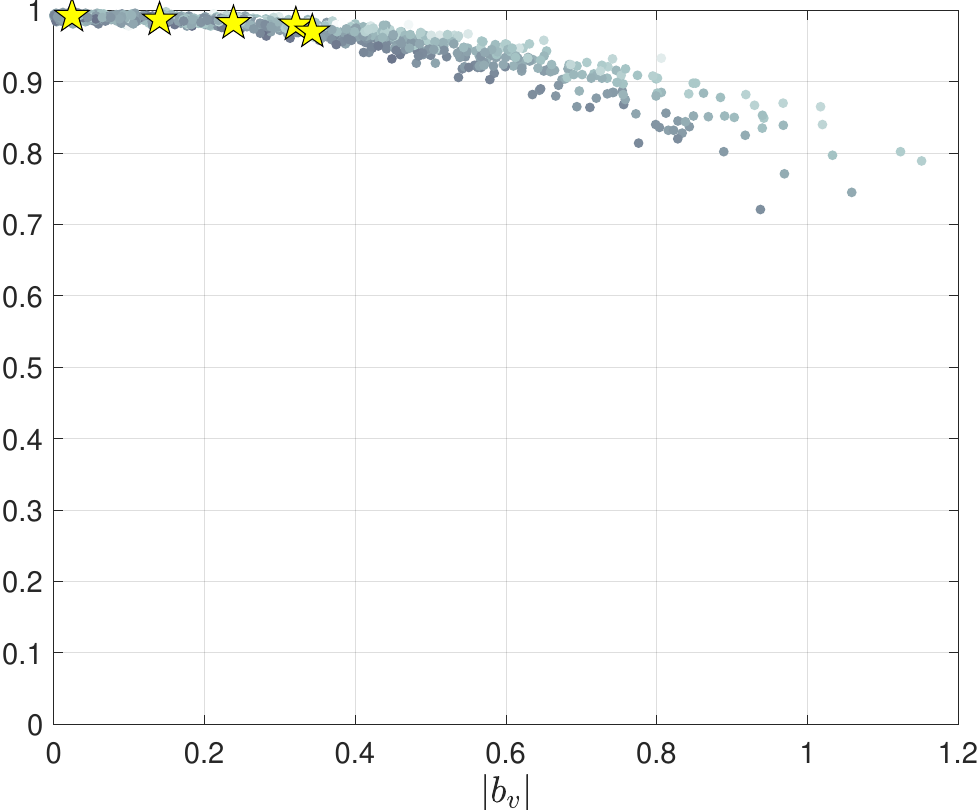}
	\end{subfigure}
	\begin{subfigure}{0.49\textwidth}
		\centering
		\includegraphics[trim=0.5in 3in 0.8in 3.3in, clip,width=\linewidth, height=0.75\linewidth]{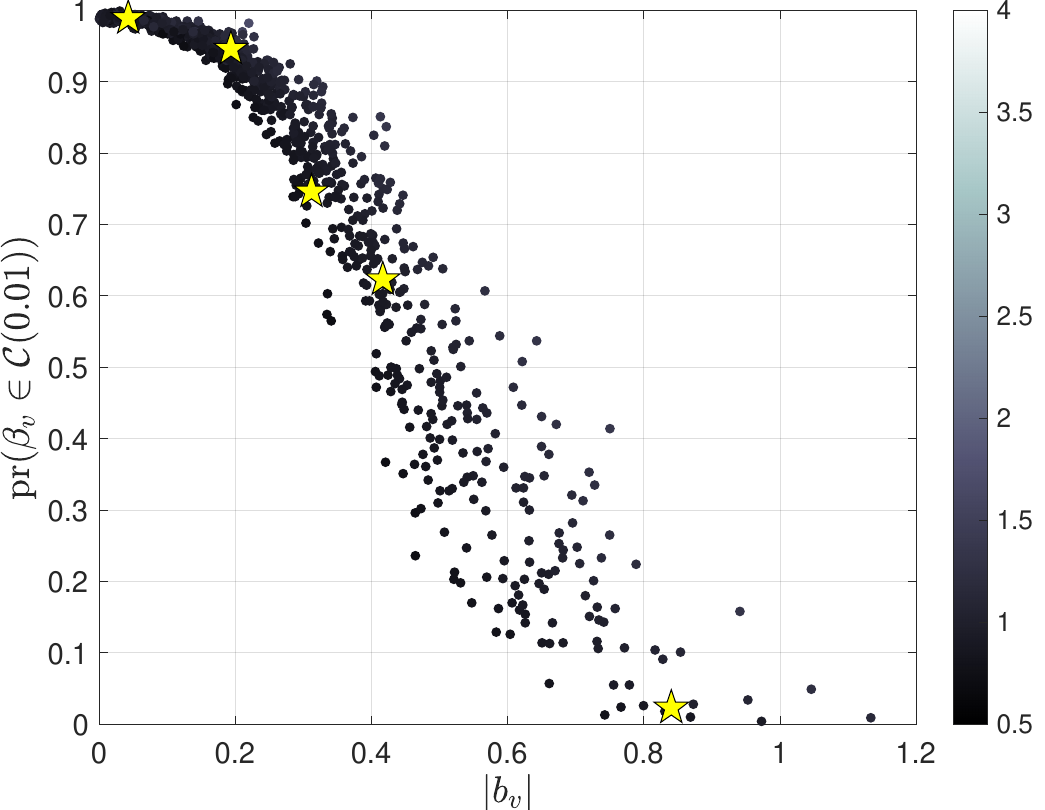} \\
	\end{subfigure}
	\caption{Coverage probabilities (vertical axis) for each $\beta_v$ plotted against the associated absolute bias $|b_{v}|$ (horizontal axis) and the mean length of the confidence interval (colour axis). High and low values of $p$ are $2450$ and $1225$ and those of $n$ are $70$ and $35$. Yellow stars correspond to signal variables. \label{figSimulations}}
\end{figure}

Detailed summary information is in Figure \ref{figSimulations}, where we plot coverage probabilities in each of the eight settings as a function of the absolute bias $|b_v|$. The $z$ axis indicated by the colour is the mean length of the intervals over simulations. Coverage probabilities for signal variables and their associated values of $|b_v|$ are represented as yellow stars. As expected in view of \S \ref{secTheory}, the coverage probabilities associated with the $\rho=0.1$ settings decay more sharply with absolute bias than with $\rho=0.9$. The decrease in $|b_v|$ for increasing $n$, as quantified in Proposition \ref{propBiasb}, is also reflected in these plots. 

Proposition \ref{propBiasb} indicates that the procedure is well calibrated asymptotically for all $\beta_v$. We confirm this empirically by taking the worst case from Table \ref{tableFactorial} ($p=2450$ and $\rho=0.1$) and considering the performance over a sequence of $n$ values.

The experiments suggest that while the procedure is valuable in settings with high correlations between columns of $X$, it is relatively less secure in the almost uncorrelated case when the sample size is small. For any given data, we suggest performing preliminary checks as in \S \ref{secNearlyReal} as an indication of the reliability of the method on the data at hand.
\begin{table}
	\begin{center}
		\begin{tabular}{c|ccc}
			& \multicolumn{3}{c}{estimated main effect} \\
	response variable  & \multicolumn{1}{c}{$\rho$} & \multicolumn{1}{c}{$n$} & \multicolumn{1}{c}{$p$}  \\
			\hline
			 modal coverage & $0.917$ & $1.000$ & $1.000$  \\
			median coverage & $6.540$ &  $1.094$ & $0.624$   \\
			median length & $1.586$  &  $\hspace{-4pt}-0.790\hspace{4pt}$ & $\hspace{-4pt}-0.015\hspace{4pt}$     \\
			\hline
		\end{tabular}
\end{center}
	\caption{Estimated main effects of $\rho$, $n$ and $p$ on the properties of our $\alpha=0.01$ nominal confidence intervals. Estimated effects of coverage are multiplicative on the odds scale. \label{tableContrasts}}
\end{table}

\begin{table}
	\begin{center}
		\begin{tabular}{c|ccccc}
			 & $n=6^2$ & $n=12^2$ & $n=18^2$ & $n=24^2$ & $n=30^2$\\
			\hline
			median coverage & 0.801 & 0.886  &  0.892 & 0.922 & 0.951 \\
			proportion $>0.9$ & 0.377 & 0.478  & 0.492 & 0.561 & 0.655 \\
			median length & 0.925  & 0.466 & 0.324 & 0.257 & 0.227 \\
			\hline
		\end{tabular}
	\end{center}
	\caption{Properties of confidence intervals for the $\rho=0.1$, $p=2450$ setting with increasing $n$. Nominal level $\alpha=0.01$. \label{tableIncreasingN}}
\end{table}

\begin{figure}
	\vspace{-1.5cm}
	\begin{center}
		$\rho = 0.9$ \hspace{6.5cm} $\rho = 0.1$ \\
	\end{center}
	\begin{subfigure}{0.01\textwidth}
		\begin{sideways}$n$ high, $p$ high
		\end{sideways}
	\end{subfigure}
	\begin{subfigure}{0.49\textwidth}
		\centering
		\includegraphics[trim=0.9in 2.9in 0.85in 3.3in, clip,width=.88\linewidth, height=0.75\linewidth]{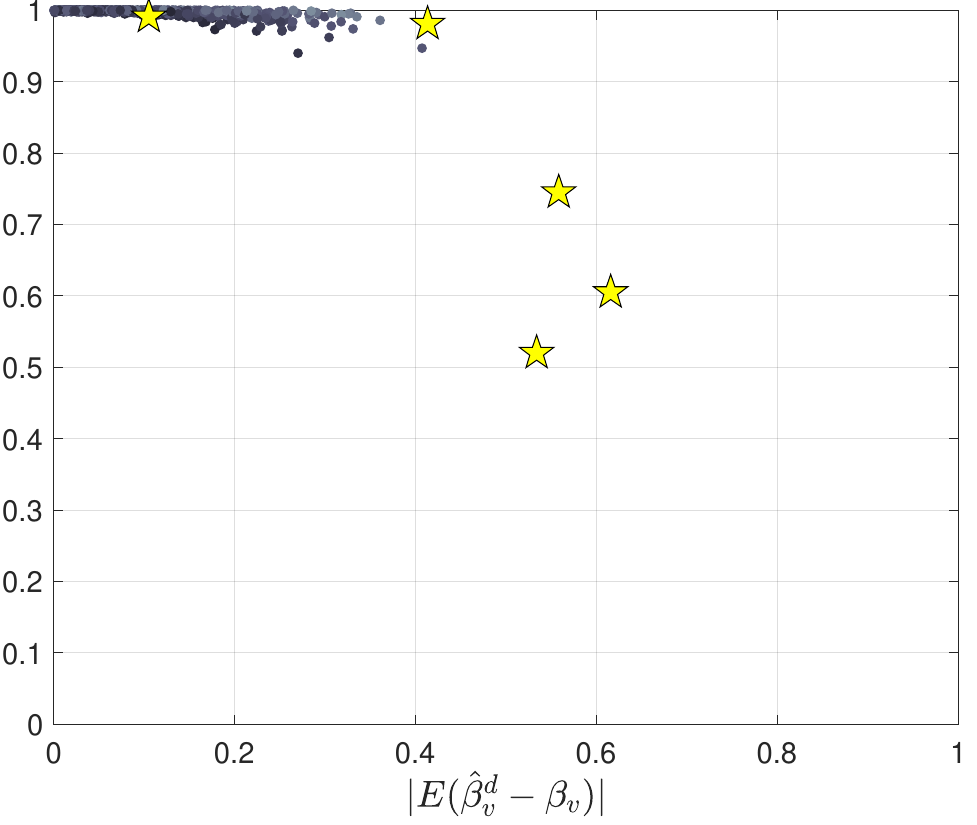}
	\end{subfigure}
	\begin{subfigure}{0.49\textwidth}
		\centering
		\includegraphics[trim=0.5in 2.9in 0.8in 3.3in, clip,width=\linewidth, height=0.75\linewidth ]{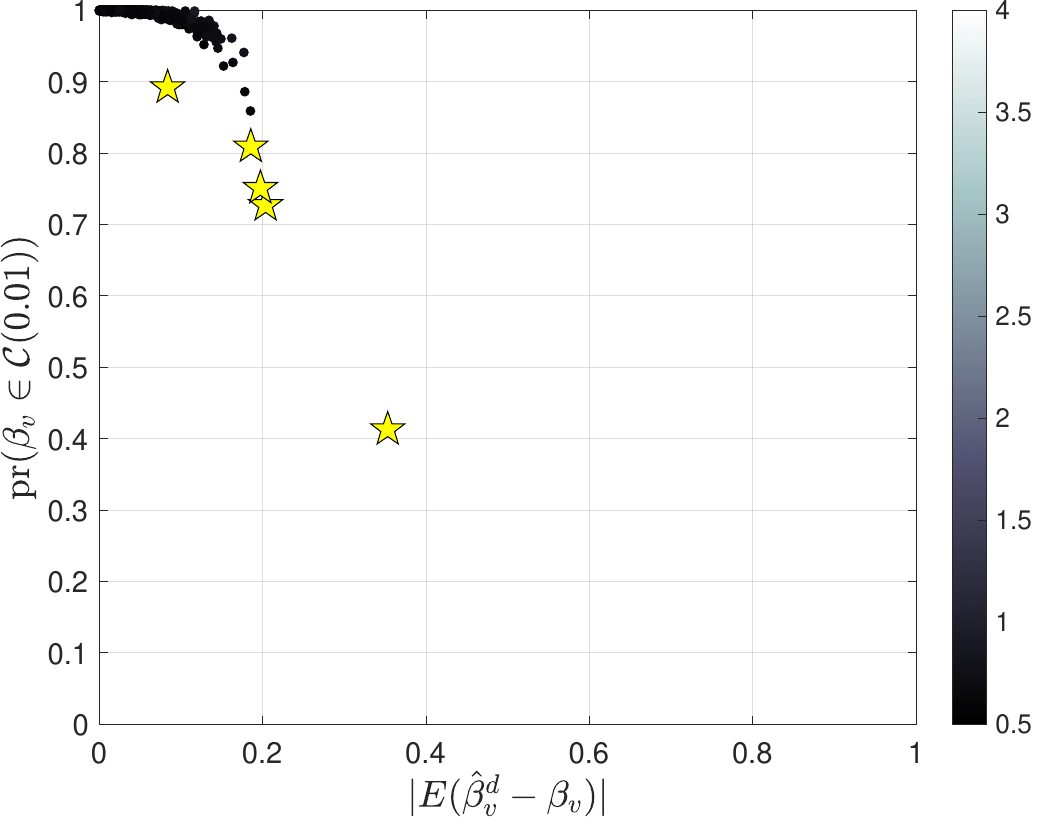} \\
	\end{subfigure}
	\begin{subfigure}{0.01\textwidth}
		\begin{sideways}$n$ high, $p$ low
		\end{sideways}
	\end{subfigure}
	\begin{subfigure}{0.49\textwidth}
		\centering
		\includegraphics[trim=0.9in 2.9in 0.85in 3.3in, clip,width=.88\linewidth, height=0.75\linewidth]{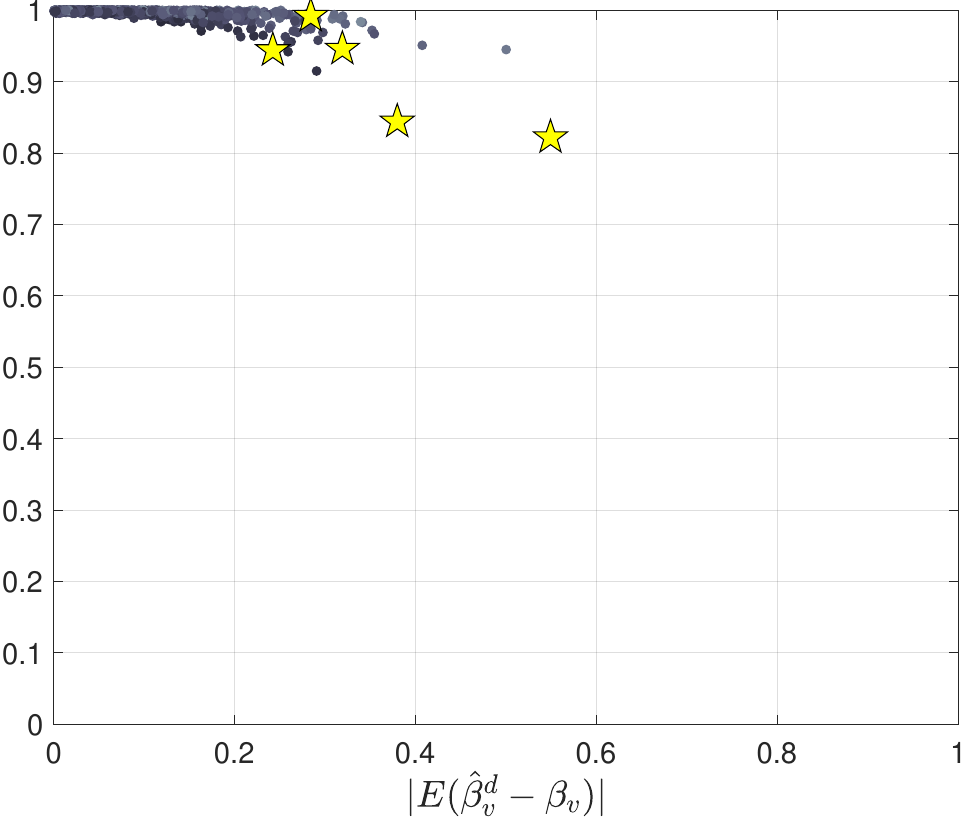}
	\end{subfigure}
	\begin{subfigure}{0.49\textwidth}
		\centering
		\includegraphics[trim=0.5in 2.9in 0.8in 3.3in, clip,width=\linewidth, height=0.75\linewidth]{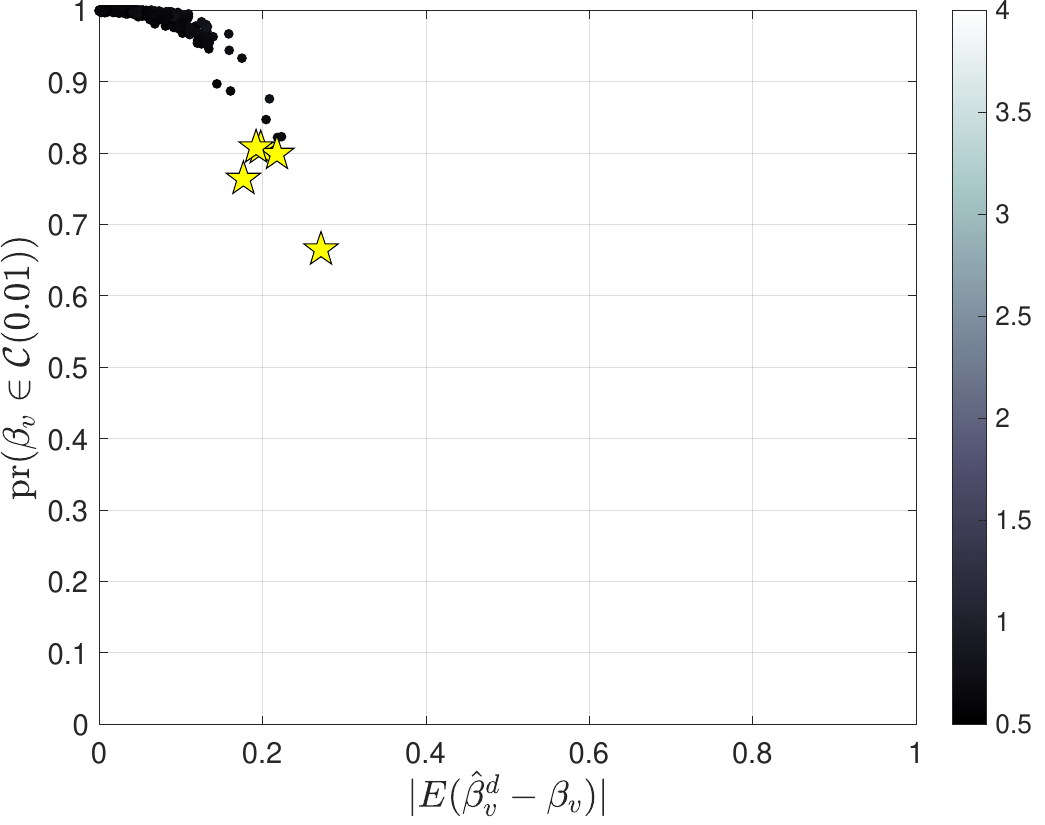} \\
	\end{subfigure}
	\begin{subfigure}{0.01\textwidth}
		\begin{sideways}$n$ low, $p$ high
		\end{sideways}
	\end{subfigure}
	\begin{subfigure}{0.49\textwidth}
		\centering
		\includegraphics[trim=0.9in 2.9in 0.85in 3.3in, clip,width=.88\linewidth, height=0.75\linewidth]{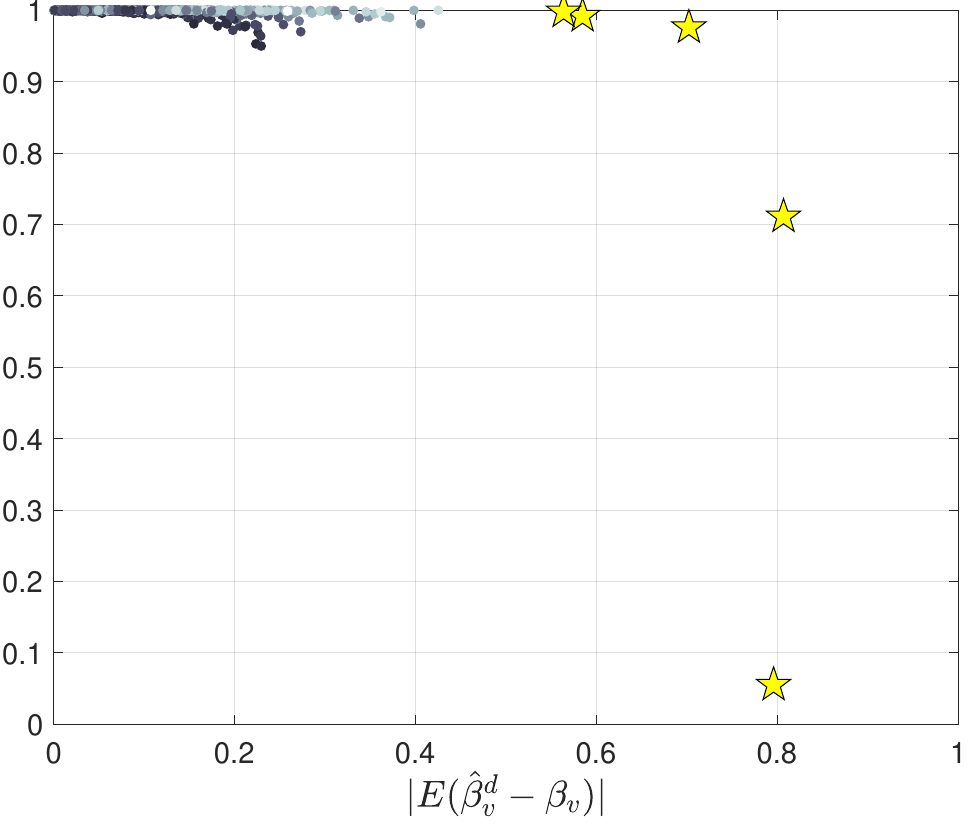}
	\end{subfigure}
	\begin{subfigure}{0.49\textwidth}
		\centering
		\includegraphics[trim=0.5in 2.9in 0.8in 3.3in, clip,width=\linewidth, height=0.75\linewidth]{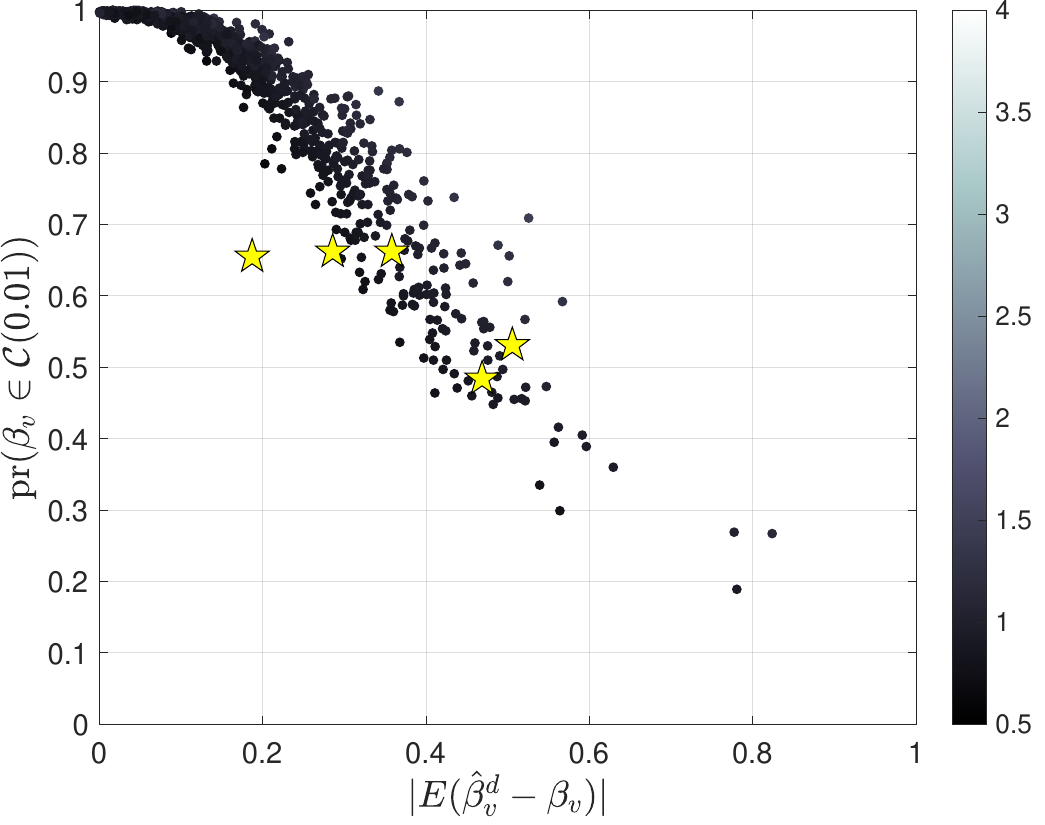} \\
	\end{subfigure}
	\begin{subfigure}{0.01\textwidth}
		\begin{sideways}$n$ low, $p$ low
		\end{sideways}
	\end{subfigure}
	\begin{subfigure}{0.49\textwidth}
		\centering
		\includegraphics[trim=0.9in 2.9in 0.85in 3.3in, clip,width=.88\linewidth, height=0.75\linewidth]{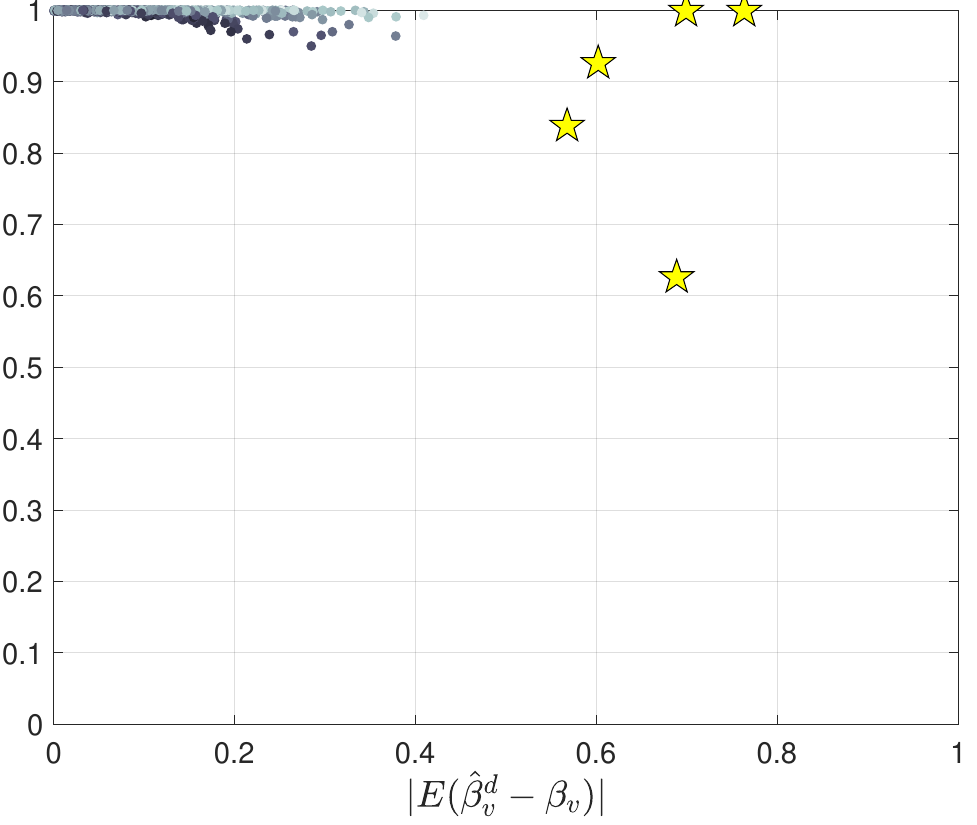}
	\end{subfigure}
	\begin{subfigure}{0.49\textwidth}
		\centering
		\includegraphics[trim=0.5in 2.9in 0.8in 3.3in, clip,width=\linewidth, height=0.75\linewidth]{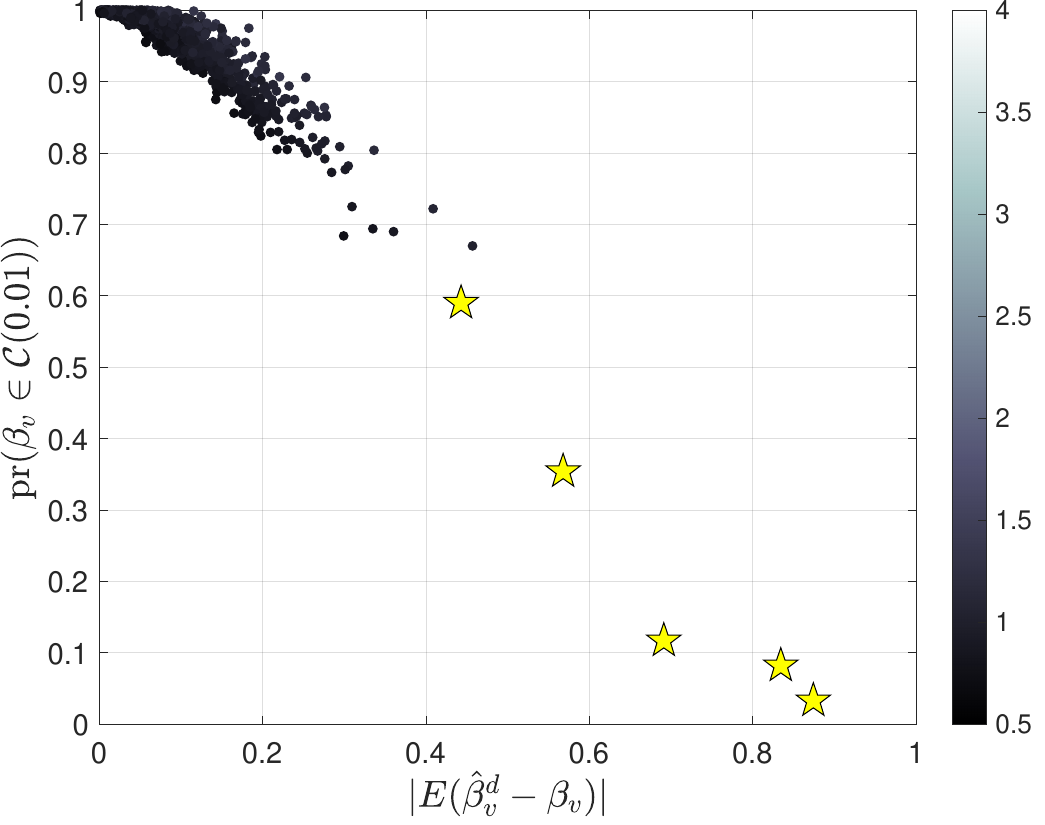} \\
	\end{subfigure}
	\caption{Coverage probabilities of debiased lasso confidence intervals (vertical axis) for each $\beta_v$ plotted against the absolute value of the Monte Carlo average of $\widehat{\beta}^{\text{d}}_v-\beta_v$ (horizontal axis) and the mean length of the confidence interval (colour axis). Yellow stars correspond to signal variables. \label{figDebiasedLasso}}
\end{figure}

Performance of the debiased lasso on the same data is summarized in Table \ref{tableDebiasedLasso} and Figure \ref{figDebiasedLasso}. Since the equicorrelation matrix used to generate $X$ does not have associated with it a sparse inverse, a key assumption for the theoretical guarantees of \cite{vdG2014} is violated. Their approach nevertheless performs favourably for variables with no real effect (see Table \ref{tableDebiasedLasso} which concerns null coefficients only), although it appears to be more systematically miscalibrated for signal variables, as illustrated in Figure \ref{figDebiasedLasso}.

\begin{table}
	\begin{center}
		\begin{tabular}{ccc|cccc}
			&     &     &	modal   & median & proportion with & median  \\
			$\rho$  & $n$ & $p$ &  coverage & coverage & coverage $>0.9$ & length  \\
			\hline			
			
0.9& 70 & 2450     & 1 & 0.999 & 1 & 1.174 \\

0.9& 70 & 1225    & 1  & 0.999 & 1 & 1.754  \\

0.9& 35 & 2450    & 1  & 1 & 1 & 2.176 \\

0.9& 35 & 1225   & 1  & 1 & 1 & 2.398  \\

0.1& 70 & 2450    & 1 & 0.999 & 0.998 & 0.647 \\

0.1&70 & 1225   & 1  & 0.999 & 0.993 & 0.641   \\

0.1&35 & 2450   & 0.998  & 0.966 & 0.682 &  0.924  \\

0.1& 35 & 1225  & 0.999 & 0.981  & 0.868 &  0.930  \\
			\hline 
		\end{tabular}
		\caption{Modal and median Monte Carlo coverage probabilities and the median of the Monte Carlo mean length of confidence intervals across the first 1000 coefficients (corresponding to null variables only) for the debiased lasso with tuning parameter chosen by cross-validation. The nominal coverage probability is $0.99$. \label{tableDebiasedLasso}}
	\end{center}
\end{table}

\section{Concluding remarks and some open problems}\label{secDiscuss}

We have proposed a new asymptotically unbiased estimator of each component of $\beta$ in a linear regression model. Our estimator has a simple closed form formula and effectively no tuning parameters, since the stability with respect to $\delta$ was found to be high. Importantly, the estimator does not require all columns of $X$ to be on the same scale. 
Theoretical conclusions associated with the procedure are confirmed by simulation in \S \ref{secFactorial}. 

We used these estimates and associated confidence intervals to propose an inferential framework based on refined confidence sets of models. The logical argument used in \S \ref{secConfModels} seems new and is of the following form. Suppose for a possible contradiction that model $m$ is true. This leads to standard least squares estimates $\hat{\beta}(m)$ of the low-dimensional vector $\beta(m)$ associated with model $m$. Under model $m$, $\hat{\beta}(m)$ is unbiased and consistent for $\beta(m)$. Let $\beta_v$ be an arbitrary entry of $\beta(m)$. To the extent that the orthogonalization makes $b_v=0$ our confidence interval for $\beta_v$ is unbiased and consistent for any model. Thus extremity of any entry of $\hat{\beta}(m)$, as calibrated by the associated limits of the corresponding confidence interval,  contradicts validity of $m$. Our intervals are asymptotically calibrated by Proposition \ref{propBiasb} but allowance for finite-sample miscalibration is discussed in \S \ref{secConfBkgrd} and \S \ref{secIllustration}.
In contrast, the usual proposed use of the debiased lasso requires adjustments  to the confidence limits  on account  of the large number of comparisons.

An alternative to the orthogonalization described in \S \ref{secCS} is to first estimate the set of signal variables, and modify \eqref{eqOptimalq} to 
\begin{equation}\label{eqOptimalq2}
q_{v}(\widehat{\mathcal{S}}) \in \biggl\{\argmin_{q\in \mathbb{R}^n} \; (q^{T}x_{v})^{-2}q^{T}(\delta I_n+\textstyle{\sum_{w\in \hat{\mathcal{S}}\backslash\{v\}}}x_{w}x_w^T)q\biggr\},
\end{equation}
where $q_v=q_{v}(\{1,\ldots,p\})$. This proposal was assessed in simulations (not reported) when $\widehat{\mathcal{S}}$ was constructed using the lasso. The coverage properties depended strongly on the lasso tuning parameter. A tuning parameter could always be found to yield a median coverage probability close to nominal. However, the coverage probabilities corresponding to the signal variables was appreciably reduced.

In \S \ref{secIllustration} we used a simple estimator of the error variance, obtained as part of the construction of calibrated confidence sets of models; see Appendix \ref{secDetailsConfSetsModels}. A similar approach, following \cite{Fan2012}, could be based on cross-validation. In factorial experiments estimates of the variance are obtained from the estimates of the null effects: if $(\varepsilon_i)_{i=1}^{n}$ are normally distributed and the approximate orthogonalization via equation \ref{eqOptimalq} was completely effective, then  
\begin{equation}\label{eqNormal2}
(q_v^{T} q_{v})^{-1}(q_{v}^{T}x_v)^2 \widetilde{\beta}_v^2
\end{equation}
provide estimates of $\tau$. The composite of $p$ such statistics is, however, contaminated by those corresponding to signal variables and by those for which the optimization problem \eqref{eqOptimalq} fails to orthogonalize $\widetilde{x}_{v}^v$ to $\widetilde{x}_{w}^v$ for one or more $w\in\mathcal{S}$. One might however construct a robust estimator of variance from something like a trimmed mean of these $p$ estimates. 

The extent to which the ideas of \S \ref{secCS} apply beyond the linear regression model remains unclear. In a generalized linear model with canonical link we can write the log-likelihood function in terms of an interest parameter $\beta_v$ as 
\begin{equation}\label{canonicalPsi}
\ell(\psi,\lambda) = \phi^{-1}\sum_{i=1}^{n}\Bigl\{Y_i (x_{i,v}\beta_v + x_{i,-v}^T \beta_{-v}) - K(x_{i,v}\beta_v + x_{i,-v}^T\beta_{-v})\Bigr\},
\end{equation}
and the function that is linear in the true parameter $\beta$ is 
\begin{equation}\label{eqGeneralGLM}
\eta_i := 	g(\mu_i)  = \{(\partial/\partial \eta)K\}^{-1}\{\mathbb{E}(Y_i)\} = x_{i}^T\beta,
\end{equation}
with $\mu_i=\mathbb{E}(Y_i)$. An estimate $\widehat{\eta}=(\widehat{\eta}_1,\ldots,\widehat{\eta}_n)^T$ 
can be expressed as
\begin{equation}\label{eqLR}
\widehat{\eta} = X\beta + \varepsilon=x_v\beta_v + X_{-v}\beta_{-v} +\varepsilon,
\end{equation}
where $\varepsilon$ captures the error in estimating $\eta$. In the linear model $\widehat{\eta}=Y$ is an estimator of $X\beta$ with error satisfying $\mathbb{E}(\varepsilon)= 0$, so that \eqref{eqLR} is a generalization. Following \S \ref{secCS} we might seek a matrix $A^v$ that approximately orthogonalizes the $v$th column, $\widetilde{x}_v^v$, of $\widetilde{X}^v=A^v X$ to its remaining columns, thereby justifying a marginal regression of $A^v \widehat{\eta}$ on $\widetilde{x}_v^v$ to provide inference on $\beta_v$. The same reparameterization arguments of \S \ref{secCS} apply. The challenge is to specify an estimator $\widehat{\eta}$, perhaps highly inefficient, but whose properties are sufficiently well understood. 

An alternative generalization might be based on the ideas of \cite{CR1987}. The possibility for this in the linear model is apparent from equation \eqref{eqPhi}, where the factor multiplying $\beta_v$ is the least squares estimate of the coefficient vector in a regression of $x_v$ on $X_{-v}$. The introduction of $\delta I_n$ would accommodate $p > n$ by replacing this least squares estimate by a ridge regression estimate \citep{HK1970} and would yield a parameter approximately orthogonal to $\beta_v$ when $p>n$. Further analysis is needed to understand how far this idea generalizes, as our preliminary calculations for a logistic regression setting (not reported) suggested some difficulties. To be useful, any subsequent operation would have to exploit the induced sparsity in the relevant column of the Fisher information matrix similarly to  \cite{NingLiu2017}, who assumed such sparsity in the inverse of this matrix without preliminary parameter-orthogonalizing steps.

\bigskip
\bigskip

\noindent \textbf{Acknowledgements.} We are grateful to Yanbo Tang and to three anonymous referees for important comments on an earlier draft. Implementation of the debiased lasso in \S \ref{secNearlyReal} and \S \ref{secFactorial} was based on highly efficient and publicly available R code written by Dr.~Rajen Shah. The work was partially supported by the UK Engineering and Physical Sciences Research Council (EP/T01864X/1) and the Natural Sciences and Engineering Council of Canada.

\begin{appendix}
	
	\section{Proofs for Section \ref{secTheory}}\label{secProofs}
	
	\subsection{Proof of Proposition \ref{propq}}\label{secProofq}
	
	\begin{proof}
		Let $M_{\delta}=(\delta I_{n} + X_{-v}X_{-v}^T)$ and write the objective function in equation \eqref{eqOptimalq} as $m=gh$, where $g(q)=(q^{T}x_v)^{-2}$ and $h(q)=q^{T}M_{\delta}q$. The associated gradient field evaluated at $q$ is $\nabla m = \nabla(g h) = g\nabla h + h \nabla g$, or more explicitly
		\begin{equation}\label{eqGradient}
		\nabla m = 2(q^T x_v)^{-2}M_{\delta}q - 2(q^T x_v)^{-3}( q^{T}M_{\delta} q) x_v.
		\end{equation}
		Any stationary point, $q_v^*$ say, solves
		\begin{equation}\label{eqToCheck}
		(q_v^{*T} x_v)M_{\delta} q_v^* = (q_v^{*T}M_{\delta} q_v^*) x_v.
		\end{equation}
		
		Since the right hand side belongs to the one-dimensional subspace of $\mathbb{R}^n$ spanned by $x_v$, we infer by inspection of the left hand side that $(\delta I_n + X_{-v}X_{-v}^{T}) q_v^*$ spans the same subspace. Equivalently, there exists an $a\in\mathbb{R}\backslash\{0\}$ such that $M_{\delta} q_v^* = a x_v$. The matrix $M_{\delta}$ is invertible for any values of $n$ and $p$ and any $\delta>0$ bounded away from zero, since the minimum eigenvalue of the Gramian matrix $X_{-v}X_{-v}^{T}$ is zero and for any eigenvalue-eigenvector pair for $X_{-v}X_{-v}^T$, $(\gamma_k,\xi_k)$ say,
		\[
		(\delta I_n+X_{-v}X_{-v}^T)\xi_k = (\delta +\gamma_k)\xi_k.
		\]
		Thus write
		\begin{equation}\label{explicitq}
		q_v^* = a (\delta I_n + X_{-v}X_{-v}^T)^{-1} x_{v}.
		\end{equation}
		Upon substitution into equation \eqref{eqToCheck} we verify that any such choice of $a$ supplies a valid solution to \eqref{eqToCheck}.
		
		This set of solutions consists of local minimizers of $m$ if and only if $\nabla \nabla^T m(q^*_v)$ is non-negative definite for any $q^*_v$ of the form given in equation \eqref{explicitq}. Equivalently $z^T\{\nabla \nabla^T m(q^*_v)\} z \geq 0$ for any $z\in\mathbb{R}^n$. Otherwise the solutions $q_v^*$ represent other critical points. Let $z\in \mathbb{R}^n$. The corresponding directional derivative, $\nabla_{z}m(q):= z^{T}(\nabla m)$, of $m$ at $q$ is
		\begin{equation}\label{eqDirectionalDeriv}
		\nabla_{z}m(q)= 2(q^T x_v)^{-2}z^{T}M_{\delta}q - 2(q^T x_v)^{-3}( q^{T}M_{\delta} q) z^T x_v.
		\end{equation}
		For $u,z \in \mathbb{R}^n$,
		\begin{eqnarray*}
			\nonumber \nabla_u\nabla_{z}m(q) = u^{T}\{\nabla (z^{T} \nabla m)\} &=& -4(q^{T}x_v)^{-3}(z^{T}M_{\delta}q) (u^{T}x_{v}) + 2(q^{T}x_v)^{-2}u^TM_{\delta}z \\
			& + & 6(q^T x_v)^{-4}(q^{T}M_{\delta}q)(z^T x_v)(u^{T}x_v) - 4(q^{T}x_v)^{-3} (u^{T}M_{\delta}q) (z^T x_v),
		\end{eqnarray*}
		so that,
		\begin{equation}\label{directionalDeriv2}
		\nabla_z\nabla_{z}m(q^*_v)=  2\{x_v^T M_{\delta}^{-1}x_v\}^{-2}z^{T}\bigl[ M_{\delta} - \{x_v^TM_{\delta}^{-1}x_v\}^{-1}x_v x_v^T \bigr] z.
		\end{equation}
		The conclusion follows because, for any function $f: \mathbb{R}^n \rightarrow \mathbb{R}$, $z^T\{\nabla \nabla^T f(q)\} z = \nabla_{z}\nabla_z f(q)$. 
		
	\end{proof}
	
	\subsection{Proof of Proposition \ref{propbias}}\label{secProofBias}
	
	\begin{proof}
		In view of the well known matrix identity
		\[
		(A+BCD)^{-1}=A^{-1}-A^{-1}B(C^{-1}+DA^{-1}B)^{-1}D A^{-1},
		\]
		we see that 
		\begin{equation}\label{eqAAProjection}
		P^v(a,\delta) = a (\delta I_n + X_{-v}X_{-v}^T)^{-1} = a \delta^{-1}(I_{n} - X_{-v}(\delta I_{p-1}+X_{-v}^{T}X_{-v})^{-1}X_{-v}^{T}).
		\end{equation}
		On setting $a=\delta$ and taking the limit as $\delta\rightarrow 0^{+}$,
		\begin{equation}\label{eqLimit}
		P^{v}=I_{n} - X_{-v}\lim_{\delta \rightarrow 0^{+}}(\delta I_{p-1}+X_{-v}^{T}X_{-v})^{-1}X_{-v}^{T} = I_{n} - X_{-v} X_{-v}^{+},
		\end{equation}
		where $X_{-v}^{+}$ is the Moore-Penrose pseudo-inverse of $X_{-v}$, the unique matrix satisfying the equations \citep{Penrose1955}:
		\begin{eqnarray}\label{eqMP}
		X_{-v} X_{-v}^{+} X_{-v} &= X_{-v} \hspace{1.5cm} 		(X_{-v} X_{-v}^{+})^{T} & = X_{-v} X_{-v}^{+} \\  
		\nonumber		X_{-v}^{+}X_{-v} X_{-v}^{+} &= X_{-v}^{+} 
		\hspace{1.5cm} 		(X_{-v}^{+} X_{-v})^{T} &= X_{-v}^+ X_{-v}. 
		\end{eqnarray}
		It follows from these equations that $I_{n}-X_{-v}X_{-v}^{+}$ is symmetric and equal to its square, from which we conclude that it is an orthogonal projection operator into the kernel of $X_{-v}^{T}$. 
	\end{proof}
	
	\subsection{Proof of Proposition \ref{propBiasb}}
	
	\begin{proof}
		Recall that $b_v= \textstyle{\sum_{w\in\mathcal{S}}}\vartheta_w \beta_w$, where $\vartheta_w=(q_v^{T}x_v)^{-1}q_v^T x_w$. When $(p-1)>n$, there exists a $(p-1)$-dimensional vector $b$, say, $(p-1)-n$ entries of which are zero, such that $x_v = X_{-v}b$. Thus, on substituting the expression \eqref{eqExpressionq} for $q_v$,
		\begin{eqnarray}\label{eqorderbias}
		\nonumber b_v &=& \{x_v^T(\delta I_{n} + X_{-v}X_{-v}^T)^{-1}x_v\}^{-1}\beta_{-v}^{T}X_{-v}^{T}(\delta I_{n} + X_{-v}X_{-v}^T)^{-1}x_v \\
		&=& \{b^{T}X_{-v}^{T}(\delta I_{n} + X_{-v}X_{-v}^T)^{-1}x_v\}^{-1}\beta_{-v}^{T}X_{-v}^{T}(\delta I_{n} + X_{-v}X_{-v}^T)^{-1}x_v.
		\end{eqnarray}
		For any $(p-1)$-dimensional vector $z$ whose entries are bounded away from zero and of order $h(n,p,\delta)$, to be discussed below, $b^T z \asymp n \cdot h(n,p,\delta)$. Similarly, by sparsity of $\beta$, $\beta^T z \asymp s \cdot h(n,p,\delta)$. Since $\delta$ is bounded away from zero, so are the entries of $X_{-v}^T\delta(\delta I_{n} + X_{-v}X_{-v}^T)^{-1}x_v=X_{-v}^T P^v(\delta,\delta)x_v$ by equation \eqref{eqGeneralP}. It follows on multiplying and dividing \eqref{eqorderbias} by $\delta$ that $b_v = O(s/n)$.
		
		It remains to consider the behaviour of 
		\begin{eqnarray*}
			\frac{q_v^T q_v}{(q_v^T x_v)^2} &=& \frac{x_v^T (\delta I_n +X_{-v}X_{-v}^T)^{-1}(\delta I_n +X_{-v}X_{-v}^T)^{-1} x_v}{(x_v^T(\delta I_n + X_{-v}X_{-v}^T)^{-1}x_v)^2} \\
			&=& \frac{\delta x_v^T \delta (\delta I_n +X_{-v}X_{-v}^T)^{-1}(\delta I_n +X_{-v}X_{-v}^T)^{-1} x_v}{(x_v^T\delta (\delta I_n + X_{-v}X_{-v}^T)^{-1}x_v)^2}.
		\end{eqnarray*}
		With $b$ the same vector as defined above, the last line of the previous equation can be written as
		\[
		\frac{\delta b^T X_{-v}^T P^v(\delta, \delta)z}{(b^T X_{-v}^T P^v(\delta,\delta)x_v)^2} \asymp \frac{\delta n h(n,p, \delta)}{(n  h(n,p,\delta))^2},
		\]
		where $z=(\delta I_n +X_{-v}X_{-v}^T)^{-1} x_v$ and the rate of convergence arises because both $X_{-v}^T P^v(\delta, \delta)z$ and $X_{-v}^T P^v(\delta, \delta)x_v$ have entries of order $h(n,p, \delta)$.	It follows that the term ignored by using $U_n(\beta_v)$ in place of $S_n(\beta_v)$ is
		\[
		\frac{(q_v^T x_v) b_v}{(\tau q_v^T q_v)^{1/2}} =O\Bigl(\frac{s}{n^{1/2}}\cdot \frac{h(n,p,\delta)^{1/2}}{\delta^{1/2}}\Bigr).
		\]
	\end{proof}

	\subsection{Proof of Proposition \ref{propAsympS}}
	
	\begin{proof}
		Let $q_{j,v}$ denote the $j$th entry of $q_v$ and write $S_{n}(\beta_v)$ as $\sum_{j=1}^n \xi_j$, where
		\[
		\xi_j := (q_{v}^Tq_v)^{-1/2} q_{j,v} \varepsilon_j.
		\]
		We have $\mathbb{E}(\xi_j)=0$ and $\text{var}(\sum_j \xi_j)=\tau$. To establish a Lindeberg-Feller central limit theorem \citep[e.g.][Theorem 4.12]{K1997} for $\sum_j \xi_j$, we need only verify the Lindeberg condition, i.e., for all $\eta>0$,
		\[
		\lim_{n\rightarrow \infty} \tau^{-1} \sum_{j=1}^n \mathbb{E}(\xi_j^2 \; \ind\{|\xi_j| > \eta \tau^{1/2}\})=0.
		\]
		Let $c_n$ denote a lower bound on $q_v^T q_v$. Note that $c_n$ is bounded away from zero because otherwise the evaluation of the objective function in \eqref{eqOptimalq} at $q_v$ is (positive and) unbounded, contradicting the definition of $q_v$ as a minimizer. Let $\zeta_n=\|q_v\|_{\max}$. Then $|\xi_j|\leq c_n^{-1}\zeta_n |\varepsilon_j|$ and the event $\{|\xi_j|>\eta \tau^{1/2}\}$ is contained in the event $\{|\varepsilon_j|>\eta \tau^{1/2}c_n \zeta_{n}^{-1}\}$. It follows that
		\begin{eqnarray*}
			\tau^{-1} \sum_{j=1}^{n} \mathbb{E}(\xi_j^2 \; \ind\{|\xi_j| > \eta \tau^{1/2}\}) &\leq & \tau^{-1} \sum_{j=1}^n \mathbb{E}(\xi_j^2 \; \ind\{|\varepsilon_j| > \eta \tau^{1/2}c_n \zeta_n^{-1}\}) \\ 
			&=&\tau^{-1} (q_v^T q_v)^{-1}\sum_{j=1}^n q_{j,v}^2 \mathbb{E}(\varepsilon_j^2 \; \ind\{|\varepsilon_j| > \eta \tau^{1/2}c_n \zeta_n^{-1}\}) \\
			&=&\tau^{-1} \mathbb{E}(\varepsilon_j^2 \; \ind\{|\varepsilon_j| > \eta \tau^{1/2}c_n \zeta_n^{-1}\}).
		\end{eqnarray*}
		Let $\delta = \eta \tau^{1/2} c_n \zeta_n^{-1}$. For any $\gamma>0$,
		\[
		\mathbb{E}(\varepsilon_j^2 \; \ind\{|\varepsilon_j|>\delta\}) \leq \mathbb{E}(\varepsilon_j^2 \delta^{-\gamma}|\varepsilon_j|^\gamma \; \ind\{|\varepsilon_j|>\delta\}) \leq \delta^{-\gamma}\mathbb{E}(|\varepsilon_j|^{2+\gamma}).
		\]
		Thus, since $\mathbb{E}(|\varepsilon_j|^{2+\gamma})<\infty$ and $\zeta_n^\gamma=\|q_
		v \|_{\max}^\gamma =o(\|q_v \|_2^\gamma) \asymp c_{n}^\gamma$,
		\[
		\lim_{n\rightarrow \infty} \tau^{-1} \sum_{j=1}^n \mathbb{E}(\xi_j^2 \; \ind\{|\xi_j| > \eta \tau^{1/2}\})=\lim_{n\rightarrow \infty}(\eta \tau^{1/2} c_n \zeta_n^{-1})^{-\gamma}\mathbb{E}(|\varepsilon_j|^{2+\gamma})=0
		\]
		and the Lindeberg condition is verified. That $\|q_
		v \|_{\max}^\gamma =o(\|q_v \|_2^\gamma)$ for any $\gamma>0$  follows from  \eqref{eqExpressionq}, which ensures that the number of non-zero elements of $q_v$ increases with $n$.
		
		Equation \eqref{eqBerryEsseen} is an application of the Berry-Esseen Theorem for non-identically distributed random variables \citep[e.g.][Chapter 5]{Petrov1995} with the constant $C$ improved by \cite{Tyurin2010}.
	\end{proof}

	\subsection{Proof of Proposition \ref{propStudentized0}}
	
	\begin{proof}
		The second term in the decomposition 
		\begin{eqnarray}\label{eqDecomp}
		\nonumber	\sup_{z\in \mathbb{R}}|\text{pr}\{V_n^{-1/2}S_n(\beta_v)\leq z\} -\Phi(z)| &\leq &	\sup_{z\in \mathbb{R}}|\text{pr}\{V_n^{-1/2}S_n(\beta_v)\leq z\} -\text{pr}\{\tau^{-1/2}S_n(\beta_v)\leq z\}| \\
		& + & \sup_{z\in \mathbb{R}}|\text{pr}\{\tau^{-1/2}S_n(\beta_v)\leq z\} -\Phi(z)| =: \text{I} + \text{II},
		\end{eqnarray}
		is bounded by $C e_n$ using Proposition \ref{propAsympS}. 
		
		For any $t>0$, define the events 
		\begin{eqnarray*}
			\mathcal{E}_1(t)&:=&\{(\tau^{-1/2}-V_n^{-1/2})S_n(\beta_v)\leq t\}, \\
			\mathcal{E}_{2}(t)&:=&\{|(V_n^{-1/2}-\tau^{-1/2})S_n (\beta_v)|\leq t\}.
		\end{eqnarray*}
		On the event $\{V_{n}^{-1/2}S_n(\beta_v)\leq z\}\cap \mathcal{E}_1(t)$,
		\[
		\tau^{-1/2} S_{n}(\beta_v)-t \leq V_n^{-1/2} S_n(\beta_v) \leq z.
		\]
		Thus $\{V_n^{-1/2}S_n(\beta_v)\leq z\}\cap \mathcal{E}_1(t)\subseteq \{\tau^{-1/2}S_n(\beta_v)\leq z + t \}$,
		and it follows on writing
		\[
		\{V_n^{-1/2}S_n(\beta_v)\} = \left\{\{V_n^{-1/2}S_n(\beta_v)\}\cap \mathcal{E}_1(t) \right\}\cup \left\{\{V_n^{-1/2}S_n(\beta_v)\}\cap \mathcal{E}_1^c(t) \right\}
		\]
		that 
		\begin{equation}\label{probUB}
		\text{pr}\{V_{n}^{-1/2}S_n(\beta_v)\leq z\} \leq \text{pr}\{\tau^{-1/2}S_n(\beta_v)\leq z+t\} + \text{pr}\{\mathcal{E}_1^c(t)\},
		\end{equation}
		where $\mathcal{E}_1^c(t)$ denotes the complement of $\mathcal{E}_1(t)$. Later $t$ will be replaced by a decreasing function of $n$, thus control over $\text{I}$ in \eqref{eqDecomp} requires an asymptotically matching lower bound on $\text{pr}\{V_{n}^{-1/2}S_n(\beta_v)\leq z\}$ as $t\rightarrow 0$. To this end, consider
		\[
		V_n^{-1/2}S_n (\beta_v) \leq \tau^{-1/2} S_n (\beta_v) + |(V_{n}^{-1/2}-\tau^{-1/2})S_n(\beta_v)|,
		\]
		so
		\[
		\{V_n^{-1/2}S_n (\beta_v)\leq z\} \supseteq \{\tau^{-1/2} S_n (\beta_v) + |(V_{n}^{-1/2}-\tau^{-1/2})S_n(\beta_v)|\leq z\} =: \mathcal{A}.
		\]
		Furthermore,
		\[
		\mathcal{A} \supseteq \mathcal{A}\cap \mathcal{E}_{2}(t) \supseteq \{\tau^{-1/2} S_n (\beta_v) \leq z-t \}\cap \mathcal{E}_2 (t),
		\]
		so that
		\begin{equation}\label{probLB}
		\text{pr}\{V_{n}^{-1/2} S_{n}(\beta_v)\} \geq  \text{pr}\{\tau^{-1/2} S_{n}(\beta_v)\leq z-t \}\text{pr}\{\mathcal{E}_{2}(t)\}
		\end{equation}
		and equations \eqref{probUB} and \eqref{probLB} together give
		\begin{eqnarray}\label{eqContinuity}
		& & 	\sup_{z\in \mathbb{R}}|\text{pr}\{V_n^{-1/2}S_n(\beta_v)\leq z\} -\text{pr}\{\tau^{-1/2}S_n(\beta_v)\leq z\}| \\
		\nonumber	 &\leq & \sup_{z':|z-z'|\leq t}\max\left\{|\text{pr}\{\tau^{-1/2}S_n(\beta_v)\leq z'\}\text{pr}\{\mathcal{E}_2(t)\} - \right. \text{pr}\{\tau^{-1/2}S_n(\beta_v)\leq z\}|, \\
		\nonumber	& &  \hspace{2.5cm} \left. |\text{pr}\{\tau^{-1/2}S_n(\beta_v)\leq z'\} - \text{pr}\{\tau^{-1/2}S_n(\beta_v)\leq z\}| + \text{pr}\{\mathcal{E}_1^c(t)\}\right\}
		\end{eqnarray}
		for any $t>0$. Additionally, by Proposition \ref{propAsympS} and  Taylor series expansion of the standard normal distribution function,
		\begin{eqnarray*}
			& & \sup_{z, z':|z-z'|\leq t}|\text{pr}\{\tau^{-1/2}S_n(\beta_v)\leq z'\} - \text{pr}\{\tau^{-1/2}S_n(\beta_v)\leq z\}| \\
			&\asymp &  \sup_{z,z':|z-z'|\leq t}|\Phi(z') - \Phi(z)| + e_n \asymp t + e_n.
		\end{eqnarray*}
		As a function of $t \rightarrow 0$, we have
		\[
		\text{pr}\{\mathcal{E}_1^c(t)\} \asymp \text{pr}\{|(V_n^{-1/2}-\tau^{-1/2})S_n (\beta_v)|\leq t\} = 1-\text{pr}\{\mathcal{E}_2(t)\}
		\]
		so that $\text{pr}\{\mathcal{E}_2(t)\}\rightarrow 1$ at the same rate in $t$ as $\text{pr}\{\mathcal{E}_1^c(t)\}\rightarrow 0$. Let $d_n$ be a decreasing function and $g_n$ a non-decreasing function, both to be specified. For two arbitrary constants $c_1$ and $c_2$, consider the event
		\[
		\mathcal{D}:=\{|(V_{n}^{-1/2}-\tau^{-1/2})||S_{n}(\beta_v)|>c_1 c_2 d_n g_n\} = \{\mathcal{D}\cap \mathcal{ B}\}\cup \{\mathcal{D}\cap \mathcal{B}^c\},
		\]
		where
		\[
		\mathcal{B}=\{|S_n(\beta_v)|\leq c_2 g_n\} = \{|\tau^{-1/2}S_n(\beta_v)|\leq \tau^{-1/2}c_2 g_n\}.
		\]
		On $\mathcal{B}$, $|S_n(\beta_v)/c_2 g_n| \leq 1$, yielding $\mathcal{D}\cap \mathcal{ B}\subseteq \{|V_{n}^{-1/2}-\tau^{-1/2}| > c_1 d_n\}$. Thus 
		\begin{equation}\label{eqE1Bound}
		\text{pr}\{|(V_{n}^{-1/2}-\tau^{-1/2})S_{n}(\beta_v)|>c_1 c_2 d_n g_n\} \leq \text{pr}(|V_{n}^{-1/2}-\tau^{-1/2}| > c_1 d_n) + \text{pr}(\mathcal{B}^c),
		\end{equation}
		where $\text{pr}(\mathcal{B}^c) \asymp \{1- \Phi(g_n)\} + e_n$ by Proposition \ref{propAsympS}. The conclusion follows by combining \eqref{eqDecomp}, \eqref{eqContinuity} and \eqref{eqE1Bound}.
	\end{proof}

	\section{Details of the analysis of \S \ref{secIllustration}}\label{secDetailsConfSetsModels}

	The cross correlations of all potential explanatory variables were examined and any {adjacent} variables whose correlations exceeded 0.95 were treated as a single variable for the purpose of constructing $\widehat{\mathcal{S}}$, with the first of each such pair being retained in the analysis. Among the 4088 variables initially present, 61 were discarded on these grounds, leaving 4027 variables.
	
	We implemented a refinement of Cox and Battey's method, as \cite{BC2018}  demonstrated that using all the available data to identify first the encompassing model $\widehat{\mathcal{S}}$ and then the confidence set of models $\mathcal{M}$ leads to lower than nominal coverage for the confidence set of models. Thus we used a sample splitting procedure, which serves to give a better calibrated set $\mathcal{M}$ but also enables unbiased estimation of error variance for construction of the confidence intervals for components of $\beta$.  The sample of size $n=71$ observations was randomly partitioned into subsets, $\mathcal{I}_1$ and $\mathcal{I}_2$, say, of sizes $n_1=35$ and $n_2 = 36$. In the terminology of \cite{CB2017}, the encompassing model was identified using  $\mathcal{I}_1$ for the cube reduction and $\mathcal{I}_2$ for the square reduction. The confidence set of models and estimate of error variance was constructed using $\mathcal{I}_1$.

	To improve the stability of this sample-splitting procedure, we repeated it 1000 times, re-randomizing the positions of the variable indices in the cube each time. The subset $\widehat{\mathcal{S}}$ was taken to be all those variables that survived the two-stage procedure in at least half of these randomizations. The confidence intervals constructed as described in \S \ref{secCS} for the 22 variables in $\widehat{\mathcal{S}}$ are reported in Table \ref{tableS}. These have been ordered according to the proportion of models from $\mathcal{M}$ to which they belong. For comparison, the lasso and the elastic net, fitted to all 71 observations and tuned to retain 22 variables, find nine and fourteen of the same variables. These are indicated in the first column of Table \ref{tableS}. 
	
	From the variables constituting the models in Table \ref{tableM2}, three were the first in a string of two or more variables whose empirical correlations exceeded 0.95 (see the first paragraph of this subsection). As far as $\mathcal{M}$ is concerned, such variables are essentially interchangeable. The affected variables are 1278 (paired with 1279), 1285 (paired with 1286, 1287, 1288 and 1289) and 4002 (paired with 4003, 4004, 4005 and 4006).

\end{appendix}

%
%


\end{document}